\pdfoutput=1
\documentclass[PAPER, atlasdraft=false, cernpreprint, UKenglish, texmf, orcidlogo]{atlasdoc}
\usepackage[full]{atlaspackage}
\usepackage{fontawesome}
\usepackage{atlasbiblatex}

\usepackage{atlasphysics}
\graphicspath{{logos/}{figures/}{aux_figures/}}

\usepackage{ANA-STDM-2024-03-PAPER-defs}

\AtlasTitle{Evidence for longitudinally polarised $Z$ bosons in electroweak $W^{\pm}Z$ production in association with two jets from $pp$ collisions at $\sqrt{s} = $ 13~TeV and 13.6~TeV with the ATLAS detector}

\AtlasAbstract{%
Measurements of polarisation states of $W$ and $Z$ gauge bosons in electroweak $W^{\pm}Z$ production in association with two jets ($WZjj\mathrm{-EW}$) in proton–proton collisions are presented.
The data set used corresponds to integrated luminosities of 140 fb$^{-1}$ and 164 fb$^{-1}$ of proton–proton collisions at centre-of-mass energies of 13 TeV and 13.6 TeV, respectively, recorded by the ATLAS detector at the CERN Large Hadron Collider.
The $W^{\pm}Z$ candidate events are reconstructed using leptonic decay modes of the gauge bosons into electrons and muons.
The observed significance for the presence in data of $WZjj\mathrm{-EW}$ events containing a longitudinally polarised  $Z$ boson is of 4.0 standard deviations.
The measured fraction of such events, integrated over the fiducial region, is $f_{\mathrm{X0}} = 0.32 \pm 0.09$, in agreement with the Standard Model prediction at next-to-leading-order in electroweak corrections.
Upper limits at 95\% confidence level of 0.19 and 0.12 are set on the fraction of events with a $W$ boson longitudinally polarised and on the fraction of events with the two bosons longitudinally polarised, respectively.
The integrated fiducial cross-section per lepton flavour of $WZjj\mathrm{-EW}$ production at $\sqrt{s} = 13.6$ TeV is also measured to be $\sigma_{WZjj\mathrm{-EW} \rightarrow \ell^{'} \nu \ell \ell jj} = 0.374 \; \pm  0.077 \,(\mathrm{stat.}) \; \pm 0.053  \,(\mathrm{syst.}) \; \pm 0.009 \,(\mathrm{lumi.})  \; \mathrm{fb}$, where $\ell$ and $\ell^{'}$ are either an electron or a muon.
This measurement is in agreement with the Standard Model expectation.
}

\AtlasRefCode{STDM-2024-03}

\PreprintIdNumber{CERN-EP-2026-213}

\arXivId{2607.20246 }

\HEPDataRecord{180113}

\AtlasJournalRef{\JHEP 11 (2026) 070}
\AtlasDOI{10.1007/JHEP11(2026)070}

\AtlasCoverEgroupAnalysisTeam{atlas-stdm-2024-03-analysis-team@cern.ch}

\hypersetup{pdftitle={ATLAS document},pdfauthor={The ATLAS Collaboration}}

\newcommand{\AtlasCoordFootnote}{%
ATLAS uses a right-handed coordinate system with its origin at the nominal interaction point (IP)
in the centre of the detector and the \(z\)-axis along the beam pipe.
The \(x\)-axis points from the IP to the centre of the LHC ring,
and the \(y\)-axis points upwards.
Cylindrical coordinates \((r,\phi)\) are used in the transverse plane,
\(\phi\) being the azimuthal angle around the \(z\)-axis.
The pseudorapidity is defined in terms of the polar angle \(\theta\) as \(\eta = -\ln \tan(\theta/2)\).
Angular distance is measured in units of \(\Delta R \equiv \sqrt{(\Delta\eta)^{2} + (\Delta\phi)^{2}}\).}

\DeclareUnicodeCharacter{306}{\v{}}

\usepackage{rotating}

\begin{document}

\maketitle

\section{Introduction}
\label{sec:intro}
%

Vector-boson scattering (VBS) is a key process to probe the electroweak (EW) symmetry breaking mechanism~\cite{Lee:1977eg,Lee:1977yc}.
The unitarity of the tree-level amplitude of the scattering of longitudinally polarised gauge bosons at high energies is governed in the Standard Model (SM) by cancellations between pure gauge-boson and Higgs-mediated contributions.
Measurements of the scattering of longitudinally polarised $W$ and $Z$ bosons therefore provide information that is complementary to direct measurements of the couplings of  the Higgs boson to gauge bosons.   %
The cross-section of the scattering of polarised $W$ and $Z$ bosons can also be modified by the presence of new-physics effects~\cite{Brass:2018hfw,Denner:2025xdz}. %

In proton--proton collisions, VBS is initiated by an interaction of two vector bosons radiated from the initial-state quarks, yielding a final state with two bosons and two jets, $VVjj$, in a purely electroweak process~\cite{Bozzi:2007ur}.
VBS diagrams are not independently gauge invariant and cannot be studied separately from other purely electroweak processes leading to the same $VVjj$ final state~\cite{Accomando:2006mc}.
At leading order in quantum chromodynamics (QCD), two categories of processes give rise to $VVjj$ final states. The first category, which includes VBS contributions, involves exclusively electroweak interactions at the Born level of order $\alpha_{\mathrm{EW}}^6$ including the boson decays, where $\alpha_{\mathrm{EW}}$ is the electroweak coupling constant.
It is referred to as EW production, labelled \vvjjew.
The second category involves both the QCD and electroweak interactions at Born level of order $\alphas^2\alpha_{\mathrm{EW}}^4$, where $\alphas$ is the strong interaction coupling constant.
It is referred to as QCD production, labelled \vvjjqcd.
According to the SM, a small interference occurs between electroweak and QCD production that also involves  both the strong and electroweak interactions at Born level of orders $\alphas\,\alpha_{\mathrm{EW}}^5$, labelled \vvjjint.
The inclusive \wzjjew production was measured at $\sqrt{s} = 13$~\TeV by the CMS and ATLAS Collaboration~\cite{CMS:2020gfh,ATLAS:2024ini} and recently at  $\sqrt{s} = 13.6$~\TeV by the CMS Collaboration~\cite{CMS:2026gqm}.

The first experimental study of polarised $VV$ scattering was performed by the CMS Collaboration in \sswwjj production~\cite{CMS:2020etf}.
The evidence for the presence of longitudinally polarised $W$ bosons in \sswwjj production was later reported by the ATLAS experiment using data collected at a centre-of-mass energy of $\sqrt{s} = 13$~\TeV~\cite{ATLAS:2025wuw}.
Experimentally, after \sswwjj production, the electroweak production of \wz in association with two jets, \wzjjew, is the second most accessible process at the Large Hadron Collider (LHC).
The first observation of joint-polarisation states of $W$ and $Z$ bosons was performed by the ATLAS Collaboration in inclusive $WZ$ diboson production~\cite{ATLAS:2022oge}, setting up general experimental techniques for polarisation measurements in diboson and VBS production.

Extending the past experimental advances from Refs.~\cite{ATLAS:2025wuw,ATLAS:2022oge} and benefiting from the increase of data brought by the Run 3 of the LHC, this paper reports on measurements of the polarisation states of the $W$ and $Z$ bosons in \wzjjew production, exploiting the fully leptonic final states where both the $Z$ and $W$ bosons decay into electrons or muons.
The $pp$ collision data were collected with the ATLAS detector at centre-of-mass energies of $\sqrt{s} = 13$~\TeV\  from 2015 to 2018, called the Run 2 period, and of $\sqrt{s} = 13.6$~\TeV\ from 2022 to 2024, referred to as the Run 3 period. The integrated luminosities recorded in the two periods corresponds to $140~\ifb$ and $164~\ifb$, respectively.
The fractions of \wzjjew events having either the $W$ or the $Z$ boson, or both simultaneously, polarised longitudinally are measured in a fiducial region close to the detector acceptance.
The polarisation vectors are defined in the \wz centre-of-mass reference frame.
The measurement phase space is dominated by \wzjjqcd events, and the expected contributions of \wzjjew events with either one of the bosons longitudinally polarised represents each only 3.6\% of all selected events.
The fraction of events with the two bosons being longitudinally polarised is only of 1\%.
The reported measurements are compared with predictions from state-of-the art Monte Carlo (MC) event generators and with the recent fixed-order theory calculation at next-to-leading-order (NLO) electroweak  accuracy~\cite{Denner:2025xdz} using \MoCaNLO~\cite{Denner:2026phn}.

%
%
%
%
%
%
%


%
\section{ATLAS detector}
\label{sec:detector}
%


The ATLAS experiment~\cite{PERF-2007-01,GENR-2019-02} at the LHC is a multipurpose particle detector
with a forward--backward symmetric cylindrical geometry and a near \(4\pi\) coverage in
solid angle.\footnote{\AtlasCoordFootnote}
It consists of an inner tracking detector surrounded by a thin superconducting solenoid
providing a \qty{2}{\tesla} axial magnetic field, electromagnetic and hadronic calorimeters, and a muon spectrometer.
The inner tracking detector (ID) covers the pseudorapidity range \(|\eta| < 2.5\).
It consists of silicon pixel, silicon microstrip, and transition radiation tracking detectors.
Lead/liquid-argon (LAr) sampling calorimeters provide electromagnetic (EM) energy measurements
with high granularity within the region \(|\eta|< 3.2\).
A steel/scintillator-tile hadronic calorimeter covers the central pseudorapidity range (\(|\eta| < 1.7\)).
The endcap and forward regions are instrumented with LAr calorimeters
for EM and hadronic energy measurements up to \(|\eta| = 4.9\).
The muon spectrometer (MS) surrounds the calorimeters and is based on
three large superconducting air-core toroidal magnets with eight coils each.
The field integral of the toroids ranges between \num{2.0} and \qty{6.0}{\tesla\metre}
across most of the detector.
The muon spectrometer includes a system of precision tracking chambers up to \(|\eta| = 2.7\) and fast detectors for triggering up to \(|\eta| = 2.4\).
The luminosity is measured mainly by the LUCID--2 detector which is located close to the beampipe.
A two-level trigger system was used to select events~\cite{TRIG-2016-01,TRIG-2022-01}.
The first-level trigger is implemented in hardware and used a subset of the detector information
to accept events at a rate close to \qty{100}{\kHz}.
This is followed by a software-based trigger that
reduced the accepted rate of complete events to \qty{1.25}{\kHz} and \qty{3}{\kHz} on average in Run~2 and Run~3, respectively,
depending on the data-taking conditions.
A software suite~\cite{SOFT-2022-02} is used in data simulation, in the reconstruction
and in the analysis of real and simulated data, in detector operations, and in the trigger and data acquisition
systems of the experiment.


%
\section{Theoretical framework and fiducial phase space}
\label{sec:framework_phase_space}
%


%
%
%
%
%
%
%
%
%
%
%
%
%
%
%
%
%
%
%
%
%
%
%
%
%
%
%
%
%
%
%
%
%
%
%
%
%
%
%
%
%
%
%
%
%
%
%
%
%
%
%
%
%
%
%
%
%
%

%
%
%
%
%

%
Polarisation is not a Lorentz invariant quantity and hence depend on a chosen reference frame.
In diboson production a natural frame choice for single-boson polarisation fraction measurements is to fix the reference $z$-axis to the direction of the corresponding boson in the diboson centre-of-mass frame~\cite{Denner:2020wz}.
This definition is used for this measurement.

In \wzjjew production, each vector boson $V$ can be polarised either longitudinally, \Vz, or transversely, \VT, leading to four distinct joint polarisation states \wzjjewzz, \wzjjewzt, \wzjjewtz and \wzjjewtt, as well as interferences between the individual polarisation states.
The respective fractions of each of the four joint polarisation states, \fzz, \fzT, \fTz and \fTT, are composed of diagonal elements of the full spin density matrix and can be interpreted as probabilities of correlated polarisation states of the two $W$ and $Z$ bosons~\cite{ATLAS:2022oge}.

The polarisation of one of the bosons can also be measured regardless of the state of the other, determining the individual polarisation fractions \fzX (\fXz) and  \fTX (\fXT) for the longitudinal and transverse polarisation states of the $W$ ($Z$) bosons, respectively.
The index X indicates either of the two 0 or T polarisation states.
The reported polarisation fractions and cross-sections are measured in a fiducial phase space chosen to closely follow  the detector-level event selection criteria described in Section~\ref{sec:selection}.
The fiducial phase space definition is based on the kinematics of particle-level objects as defined in Ref.~\cite{TruthWS}.
These are final-state prompt\footnote{A prompt lepton is a lepton that is not produced in the decay of a hadron, a $\tau$-lepton, or their descendants.} leptons associated with the $W$ and $Z$ boson decays.
Particle-level charged leptons after quantum electrodynamics final-state radiation are ``dressed''
by adding to the lepton four-momentum the contributions from photons with an angular distance $\Delta R \equiv \sqrt{(\Delta\eta)^2 + (\Delta\phi)^2} < 0.1$
from the lepton. Dressed leptons, and final-state neutrinos that do not originate from hadron or $\tau$-lepton decays, are matched to the $W$ and $Z$ boson decay products using %
an algorithm that does not depend on details of the MC generator,
called the ``resonant shape'' algorithm~\cite{ATLASWZ_8TeV}.
The dressed charged leptons from $Z$ and $W$ boson decay are required to have $|\eta| < 2.5$.
Transverse momenta \pt\ of leptons from the $Z$ boson are required to be above $15$~\GeV.
Transverse momentum of the charged lepton from the $W$ boson is required to be above  $20$~\GeV.
The invariant mass of the two leptons from the $Z$ boson decay must differ by at most $10$~\GeV{}
from the world average value of the $Z$ boson mass $m_Z^{\textrm{PDG}}= 91.1876$~\GeV~\cite{Tanabashi:2018oca}. %
The $W$ transverse mass, defined as $m_{\textrm{T}}^{W} = \sqrt{2 \cdot p_{\mathrm{T}}^\nu \cdot p_{\mathrm{T}}^\ell \cdot[1 -\cos{\Delta \phi(\ell, \nu)}]}$, where $\pt^\ell$ and $\pt^\nu$ are the transverse momenta of the lepton from $W$ boson decay and of the neutrino, respectively, and $\Delta \phi(\ell, \nu)$
is the angle between them in the transverse plane, must be greater than $30$~\GeV.
In addition, it is required that the angular distance $\Delta R$
between the charged leptons from the $W$ and $Z$ decay is larger than $0.3$, and that
$\Delta R$ between the two leptons from the $Z$ decay is larger than $0.2$.

In addition to these requirements, at least two jets with $\pt>40$~\GeV\ and $|\eta_j|< 4.5$ are required.
These particle-level jets are reconstructed from stable particles with a lifetime of $\tau>30$~ps in the simulation after parton showering, hadronisation, and decay of particles with $\tau<30$~ps.
Muons, electrons, neutrinos and photons associated with $W$ and $Z$ decays are excluded.
The particle-level jets are reconstructed using the anti-$k_{t}$~\cite{antikt,Fastjet} algorithm with a radius parameter $R$ = $0.4$.
The angular distance between all selected leptons and jets is required to be $\Delta R (j,\ell)>0.3$.
If this requirement is not satisfied, the jet is discarded.
The two highest-\pT jets in opposite hemispheres of the detector coordinate system, $\eta_{j1} \cdot \eta_{j2} < 0$, are selected.
These two jets are referred to as tagging jets.
Their invariant mass, $m_{jj}$, is required to be $m_{jj}>500$~\GeV.
Finally, hard processes with a $b$-quark in the initial or final states at the matrix-element level, such as $tZj$ production,  are not considered as part of the \wzjjew events.
The production of $tZj$ results from a $t$-channel exchange of a $W$ boson  between a $b$- and a $u$-quark giving a final state with a $t$-quark, a $Z$ boson and a light-quark jet, but does not include diagrams with gauge boson couplings.

The contribution from off-shell effects is minimised by the tight invariant mass window of $\pm 10$~\GeV\ around the nominal $Z$ boson mass used for the definition of the fiducial phase space.
This contribution is estimated from NLO EW calculations to amount to $1.5$\% at the integrated level and in the shape of most of the differential distributions~\cite{Denner:2025xdz}.
The small non-resonant contributions present in the fiducial phase space are included in the measured events and therefore in the defintion of the measured polarisation fractions.
Interferences among polarisation states are also expected to be small, amounting to $2$\% according to NLO EW calculations~\cite{Denner:2025xdz}.
Their contribution is  therefore not considered separately, but is counted in the measured polarisation fractions: the total \wzjjew cross-section is defined as the sum of the polarised contributions.

%
%
%

%

%
%
%
%


%
\section{Signal and background simulation}
\label{sec:samples}
%


%

%
%
%
The polarised on-shell \wzjjewzz, \wzjjewzt, \wzjjewtz, and \wzjjewtt signal processes were simulated in the narrow-width approximation using \SHERPA[3]~\cite{Sherpa:2024mfk,Hoppe:2023uux} with up to one additional parton at leading-order (\LO) in QCD and EW couplings.
The \LO-accurate matrix elements were matched to the \Sherpa parton
shower (PS) based on Catani--Seymour dipole factorisation~\cite{Gleisberg:2008fv,Schumann:2007mg} using the MEPS@LO
prescription~\cite{Hoeche:2011fd,Hoeche:2012yf,Catani:2001cc,Hoeche:2009rj}.
This multijet merging approach, labelled 0,1j@LO, was shown to capture the bulk of \NLO QCD corrections to polarisation fractions, as they typically stem from real emission rather than virtual diagrams~\cite{Hoppe:2023uux,Pelliccioli:2023zpd}.
Divergencies coming from triboson contributions are avoided by keeping vector boson masses finite. Possible gauge dependencies are verified to be small by comparing to a full off-shell prediction from \SHERPA[3] at the same perturbative order.
Diagrams with a $b$-quark in either the initial or final state were not considered.
The \PDFforLHC[21]~\cite{PDF4LHCWorkingGroup:2022cjn} set of parton distribution function (PDF) was used. %
This MC simulation provides the most accurate particle-level prediction available for these signal processes.

For the estimate of modelling uncertainties, alternative polarised MC samples were generated at \LO using the \MGNLO[2.9.16]~\cite{BuarqueFranzosi:2019boy} (3.5.7 for Run 3) MC event generator, the \NNPDF[3.0nlo]~\cite{Ball:2014uwa} PDF set, and interfaced with \PYTHIA[8.307p1]~\cite{Sjostrand:2014zea} (8.308 for Run 3) using the A14 tune~\cite{ATL-PHYS-PUB-2014-021} for the modelling of the PS, hadronisation and underlying event.
The four-flavour scheme was used for the generation.
Inclusive production of \wzjjew events was simulated using the \SHERPA[2.2.12] (2.2.14 for Run 3) MC event generator~\cite{Bothmann:2019yzt} using the \NNPDF[3.0nnlo] PDF set.
Processes of order six in $\alpha_\mathrm{EW}$ and zero in $\alphas$ with up to one additional parton at \LO in QCD and EW were included and merged with the \SHERPA parton shower.
Diagrams with a $b$-quark in either the initial or final state are not considered.
The \LO-accurate matrix elements for up to one additional parton were matched to a PS based on Catani--Seymour dipole factorisation~\cite{Gleisberg:2008fv,Schumann:2007mg} using the MEPS@LO prescription~\cite{Hoeche:2011fd,Hoeche:2012yf,Catani:2001cc,Hoeche:2009rj}.

The \wzjjqcd contribution was also simulated using \SHERPA[2.2.12] (2.2.16 for Run 3).
For \wzjjqcd processes, the matrix elements contain all diagrams with four electroweak vertices.
They were calculated for four leptons and up to one other parton at \NLO in QCD and up to three other partons at \LO using \textsc{Comix}~\cite{Gleisberg:2008fv} and \OPENLOOPS~\cite{Cascioli:2011va}, and merged with the \Sherpa PS~\cite{Schumann:2007mg} according to the ME+PS$@$NLO prescription~\cite{Hoeche:2012yf}.
The \NNPDF[3.0nnlo] PDF set was used.
A second MC sample for \wzjjqcd production, used to estimate modelling uncertainties, was simulated using the \MGNLO[2.6.5] (3.5.7 for Run 3) MC event generator, interfaced with \PYTHIA[8.240] (8.308 for Run 3) using the A14 set of tuned parameters~\cite{ATL-PHYS-PUB-2014-021} for the modelling of the PS, hadronisation and underlying event.
Matrix elements containing three leptons, one neutrino and up to two partons in the final state were calculated at \NLO in QCD and merged with the PS from  \PYTHIA[8.210] using the \FXFX scheme~\cite{Frederix:2012ps}.
The \NNPDF[3.0nlo] PDF set was used for the hard-scattering process,  while the \NNPDF[2.3lo]~\cite{Ball:2012cx}  PDF set was used for the PS.
The default dynamic renormalisation and factorisation scales set by \MGNLO~\cite{Hirschi:2015iia} were used.

The  contribution from interference between \wzjjew and \wzjjqcd processes, labelled \wzjjint, was simulated at \LO using \MGNLO[2.6.5]~\cite{Alwall:2014hca} (3.5.7 for Run 3), the \NNPDF[3.0nlo] PDF set, and interfaced with \PYTHIA[8.240]~\cite{Sjostrand:2014zea} (8.308 for Run 3) using the A14 tune~\cite{ATL-PHYS-PUB-2014-021} for the modelling of the PS, hadronisation and underlying event.

The background sources not originating from \wzjj production include processes with two or more electroweak gauge bosons, namely $ZZ$, $WW$ and $VVV$ ($V=W,Z$); processes with top quarks, such as \tt, \ttV, single top and \tzj; and processes with gauge bosons associated with jets or photons ($Z+j$ and $Z\gamma$).
MC simulation is used to estimate the contribution from  background processes with three or more prompt leptons.
Background processes with at least one misidentified lepton are evaluated using data-driven techniques and simulated events are used to assess the systematic uncertainties of these backgrounds.

The \SHERPA[2.2.2] (2.2.16 for Run 3) event generator was used to simulate both the $q\bar{q}$ and $gg$-initiated $ZZ$, including $H \rightarrow VV$ production, using the \NNPDF[3.0nnlo] PDF set.
It provides a matrix element calculation accurate at \NLO in $\alpha_{\mathrm{s}}$ for 0- and 1-jet final states, and \LO accuracy for 2- and 3-jet final states.
The \zzjjew production at $\sqrt{s} =$~13~\TeV was modelled  using the \MGNLO[2.6.7] generator with matrix elements calculated at \LO in QCD and with the \NNPDF[3.0nlo] PDF set.
The events were interfaced with \PYTHIA[8.230] using the A14 tune and the \NNPDF[2.3lo] PDF set.
For $\sqrt{s} =$~13.6~\TeV, the \zzjjew production was simulated using the \SHERPA[2.2.14] event generator and the same parameters as used for \wzjjew production.

The production of \ttV\ events was modelled using the \MGNLO[2.3.3] (3.5.1 for Run 3) generator at \NLO with the \NNPDF[3.0nlo] PDF.
The events were interfaced to \PYTHIA[8.210] (8.309 for Run 3) using the A14 set of tuned parameters and the \NNPDF[2.3lo] PDF set.
The \tzj process was modelled using the \MGNLO[2.3.3] (3.5.1 for Run 3) generator at \NLO with the \NNPDF[3.0nlo] PDF.
The events were interfaced with \PYTHIA[8.230] (8.309 for Run 3) using the A14 set of tuned parameters and the \NNPDF[2.3lo]  PDF set.
Finally, the on-shell production of triboson events with fully leptonic decays were simulated by the \SHERPA[2.2.2] (2.2.14 for Run 3) event generator using factorised gauge-boson decays.
Matrix elements, accurate to \NLO for the inclusive process and to \LO for up to
two additional parton emissions, were matched and merged with the \SHERPA parton
shower based on Catani--Seymour dipole factorisation using the \MEPSatNLO prescription.
Samples were simulated using the \NNPDF[3.0nnlo] PDF set.

For all simulated processes relying on \Pythia for the hadronisation, the decays of bottom and charm hadrons were simulated using the \EVTGEN[1.2.0] program~\cite{Lange:2001uf}.
All generated MC events were passed through the ATLAS detector simulation~\cite{Aad:2010ah}, based on
\GEANT~\cite{Agostinelli:2002hh}, and processed using the same reconstruction software as used for the data.
For MC events generated at $\sqrt{s} = 13$~\TeV, multiple interactions in the same and neighbouring bunch crossings (pile-up) were modelled by overlaying the hard-scattering event with simulated inelastic $pp$ events generated with \PYTHIA[8.186] using the \NNPDF[2.3lo] PDF set and the A3~\cite{ATL-PHYS-PUB-2016-017} set of tuned parameters. %
For MC events generated at $\sqrt{s} = 13.6$~\TeV, the effect of \pileup
was modelled by overlaying~\cite{SIMU-2020-01} the simulated hard-scattering event
with inelastic proton--proton (\(pp\)) events generated from a mix of
\EPOS[2.0.1.4]~\cite{Werner:2005jf} and \PYTHIA[8.308]~\cite{Bierlich:2022pfr}.
The \EPOS events were generated with the \EPOS LHC tune~\cite{Pierog:2013ria}
and the \PYTHIA events with the A3 tune~\cite{ATL-PHYS-PUB-2016-017}
and the \NNPDF[2.3lo]~\cite{Ball:2012cx} set of parton distribution functions (PDF).
\PYTHIA pileup events include either a high transverse momentum (\pt) jet,
a prompt photon, or a lepton from a $b$-hadron decay,
while \EPOS was filtered to simulate all remaining pileup events in the overlay sample.
The individual simulations were reweighted to ensure a smooth connection across jet \pt.
The MC events were weighted to reproduce the distribution of the average number of interactions per bunch crossing
(\(\left<\mu \right>\)) observed in the data~\cite{Aaboud:2016mmw}.
Furthermore, the lepton and jet momentum scale and resolution, the lepton reconstruction, identification, isolation and trigger efficiencies and the pile-up jets and $b$-jet selection efficiencies in the simulation were corrected to match those measured in data.

%
%
%
%
%
%


%
\section{Event reconstruction and selection}
\label{sec:selection}
%


Only data recorded with stable beam conditions and with all relevant detector subsystems operational are considered~\cite{DAPR-2018-01}.
Candidate events are selected using triggers requiring at least one electron or muon~\cite{TRIG-2016-01,ATLAS:2019dpa,ATLAS:2020gty}.
These triggers require leptons to satisfy transverse momentum thresholds and isolation criteria that depend on the data-taking run period and the instantaneous luminosity.
The primary vertex is defined as the reconstructed vertex with at least two charged-particle tracks that has the largest sum of the $\pT^{2}$ for the associated tracks.
Events are required to have a primary vertex compatible with the LHC luminous region inside the ATLAS detector.

Muon candidates are identified by tracks reconstructed in the MS and matched to tracks reconstructed in the ID.
Muons are required to satisfy a ``medium'' identification selection~\cite{ATLAS:2020auj}.
The muon momentum is measured by combining the MS measurement, corrected for the energy deposited in the calorimeters, and the ID measurement.
The $\pt$ of the muon must be greater than $15$~\GeV{} and its pseudorapidity must satisfy $|\eta|<2.5$.

Electron candidates are reconstructed from energy clusters in the electromagnetic calorimeter matched to ID tracks.
Electrons are identified using a discriminant that is the value of a likelihood function constructed with information about the shape of the electromagnetic showers in the calorimeter, the track properties and the quality of the  track-to-cluster matching for the candidate~\cite{ATLAS:2019qmc, ATLAS:2023dxj}.
Electrons must satisfy a ``medium'' likelihood requirement. %
The electron momentum is computed from the cluster energy and the direction of the track.
The $\pt$ of the electron must be greater than $15$~\GeV{} and the pseudorapidity of the cluster must satisfy
$\abseta< 1.37$ or $1.52 < \abseta < 2.47$, to avoid the transition region between the cryostats.

Electron and muon candidates are required to originate from the primary vertex.
Thus, the significance of the track's transverse impact parameter calculated relative to the beam line, $|{d_{0}/\sigma_{d_{0}}}|$, must be smaller than $3.0$ for muons and less than $5.0$ for electrons.
Furthermore, the longitudinal impact parameter, $z_{0}$ (the difference between the value of $z$ of the point on the track at which $d_{0}$ is defined and the longitudinal position of the primary vertex), is required to satisfy $|z_0\cdot \sin(\theta)|<0.5$~mm.
Electrons and muons are required to be isolated from other particles using both calorimeter-cluster and ID-track information~\cite{ATLAS:2020auj,ATLAS:2019qmc, ATLAS:2023dxj}.

Jets of hadrons are reconstructed using a particle-flow algorithm based on noise-suppressed positive-energy topological clusters in the calorimeters~\cite{ATLAS:2017ghe}.
Energy deposited in the calorimeters by charged particles is subtracted and replaced by the momenta of tracks matched to those topological clusters.
Tracks which are not matched to the primary vertex are not used in jet reconstruction, which effectively removes contribution from charged particle pile-up.
The jets are clustered using the anti-$k_{t}$ algorithm with a radius parameter $R = 0.4$.
They are  calibrated according to {\textit{in situ}} measurements of the jet energy scale~\cite{JETM-2018-05}.
All jets must have $\pt>25$~\GeV{} and be reconstructed in the pseudorapidity range $|\eta|<4.5$.
To reduce the impact of jets originating from pile-up interactions, jets with $|\eta|<2.4$ and $\pt < 60$~\GeV, or with $2.5 < |\eta| < 4.5$ and $\pt < 120$~\GeV, are required to satisfy the ``tight'' and ``loose'' working points of the jet vertex~\cite{Aad:2015ina} and forward jet vertex~\cite{ATL-PHYS-PUB-2019-026} tagging algorithms, respectively.
Jets with $|\eta|<2.5$ containing a $b$-hadron are identified in Run~2 data and simulation with a deep-learning neural network (NN), DL1r~\cite{FTAG-2019-07}, which uses distinctive features of the $b$-hadron decay, like impact parameters of the tracks and displaced vertices reconstructed in the inner detector.
Jets initiated by $b$-quarks are selected by setting the algorithm's output threshold such that a $70\%$ $b$-tagged jet selection efficiency is achieved in simulated $t\bar{t}$ events, with a rejection factor of about $600$ against light-flavour jets~\cite{FTAG-2019-07,FTAG-2019-02,FTAG-2020-08,FTAG-2018-01}.
In Run~3 data and simulation, a machine-learning algorithm, GN2,  based on a transformer architecture using tracks and jet kinematic variables as inputs to the training is used~\cite{ATLAS:2025dkv}.
The threshold on the algorithm's output discriminant is chosen such that about $70\%$ of $b$-tagged jets in simulated \tt events have a discriminant score above that threshold. The rejection factor of light-flavour jets is about 1700~\cite{ATLAS:2025dkv,ATLAS:2026bvi,ATLAS:2026hwn}.
Discrepancies of $b$-jet efficiencies and light and charm mistag rates between data and simulation are corrected with dedicated calibration scale factors~\cite{FTAG-2019-02, FTAG-2020-08,FTAG-2018-01,ATLAS:2026bvi,ATLAS:2026hwn}.

To avoid cases where the detector response to a single physical object is reconstructed as multiple different final-state objects,  e.g. an electron reconstructed as both an electron and a jet, several steps are followed to remove such overlaps, as described in Ref.~\cite{Aaboud:2016lpj}.

The transverse momentum of the neutrino is estimated from the missing transverse momentum in the event, \met,
calculated as the negative vector sum of the transverse momentum of all identified hard physics objects
(electrons, muons, jets), with a contribution from an additional soft term.
This soft term is calculated from ID tracks matched to the primary vertex and not assigned to any of the hard objects~\cite{JETM-2020-03}.

Events are required to contain exactly three lepton candidates satisfying the selection criteria described above.
To ensure that the trigger efficiency is well determined, at least one of the candidate leptons is required to have
$\pt > 25$~\GeV\ for the year $2015$ and $\pt > 27$~\GeV\ for other running periods, as well as being  geometrically matched to a lepton that was selected by the trigger.

To suppress background processes with at least four prompt leptons, events with a fourth lepton candidate
satisfying looser selection criteria are rejected. For this looser selection, the lepton $\pt$ requirement is
lowered to $\pt>5$~\GeV{}, electrons are allowed to be reconstructed in the region $1.37< \abseta< 1.52$ and muons in the range of $\abseta< 2.7$. The ``loose'' identification requirements~\cite{ATLAS:2019qmc,ATLAS:2020auj} are used for both the electrons and muons.
A less stringent requirement is applied for electron isolation and is based only on ID track information.
No dedicated identification algorithm is used to suppress events with electrons and muons originating from the decay of $\tau$-leptons.

Candidate events are required to have at least one pair of leptons with the same flavour and opposite charge,
with an invariant mass that is consistent with $m_Z^{\textrm{PDG}}$ to within $10$~\GeV.
This pair is considered to be the $Z$ boson candidate. If more than one pair can be formed, the pair whose invariant mass is closest to $m_Z^{\textrm{PDG}}$ is taken as the \Zboson\ boson candidate.
The remaining third lepton is assigned to the $W$ boson decay. %
The transverse mass of the $W$ candidate, computed using \met and the \pt of the associated lepton, is required to be greater than $30$~\GeV.

Backgrounds originating from misidentified leptons are suppressed by requiring the lepton associated with the $W$ boson to satisfy more stringent selection criteria.
The transverse momentum of this lepton is therefore required to be greater than $20$~\GeV.
This lepton is also required to satisfy ``tight''  identification and isolation requirements. %

To select \wzjj\ candidates, events are further required to be associated with at least two ``tagging'' jets.
The leading tagging jet is selected as the highest-\pt\ jet in the event with $\pt > 40$~\GeV.
The second tagging jet is selected as the one with the highest \pt\ among the remaining jets that have a pseudorapidity of opposite sign to the first tagging jet and a $\pt > 40$~\GeV.
These two jets are required to satisfy $m_{jj} > 150$~\GeV, to minimise the contamination from triboson processes.
This selection phase space is used to define background control regions.
The final signal region (SR) for \wzjjew processes is defined by requiring that the invariant mass of the two tagging jets, $m_{jj}$, be greater than $500$~\GeV\ and that no $b$-tagged jet be present in the event.


%
\section{Background estimation}
\label{sec:background}
%

The background sources are classified into two groups:  events where all lepton candidates are prompt leptons or are produced in the decay of a $\tau$-lepton (irreducible background) and events where at least one of the candidate leptons is not a prompt lepton (reducible background). Candidates that are not prompt leptons are also called  ``misidentified'' or ``fake'' leptons.

The dominant source of background originates from the QCD-induced production of \wz dibosons in association with two jets, \wzjjqcd.
The shapes of distributions of kinematic observables of this irreducible background are modelled by the \sherpa MC simulation.
The other main sources of irreducible background arise from $ZZ$ and $t\bar{t} + V$ (where $V$ = $Z$ or $W$).
These irreducible backgrounds are modelled using MC simulations.
Data in two dedicated control regions (CR), referred to as $ZZ$-CR and $b$-CR, respectively, are used to constrain the normalisations of the \zzjjqcd,  \ttV and \tzj backgrounds.
The control region $ZZ$-CR, enriched in $ZZ$ events, is defined by applying the \wzjj event selection defined in Section~\ref{sec:selection}, but instead of vetoing a fourth lepton, events must have at least a fourth lepton candidate with looser identification requirements.
This region is dominated by \zzjjqcd events with a contribution of \zzjjew\ events of the order of 12\%.
The control region $b$-CR, enriched in \ttV and \tzj events, is defined by selecting \wzjj\ candidate events having at least one reconstructed $b$-tagged jet.
This region also contains contributions from \wzjjqcd and reducible background events.
The remaining sources of irreducible background are events from \wzjjew production with at least one $\tau$-lepton decay, which are also considered as backgrounds and not part of the measured signal, \wzjjint , \zzjjew and $VVV$ events.
Contributions of these processes in the signal region are lower than 1\% and are estimated from MC simulations.

The reducible backgrounds originate from $Z+j$, $Z\gamma$, $t\bar{t}$, $Wt$ and $WW$ production processes.
These backgrounds are estimated differentially by using a data-driven method, the matrix method,  based on the inversion of a global matrix containing the efficiencies and the misidentification probabilities for prompt and fake leptons~\cite{ATLASWZ_8TeV,Aad:2014pda,WZPol36fb,ATLAS:2022swp,ATLAS:2024ini}.
Another independent method of assessing the reducible background was also considered.
This method estimates the amount of reducible background using MC simulations scaled to data by process-dependent factors determined from the data-to-MC comparison in dedicated control regions.
Agreement within $20\%$ with the matrix method estimate is obtained in both yield and shape of the distributions of reducible background events.
This level of agreement is compatible with the systematic uncertainties determined for the matrix method estimate.

The number of events expected from MC simulation and data-driven background estimates are summarised in Table~\ref{tab:EventYields_prefit} and compared with the number of observed data events, for the signal region and the two control regions.
All sources of uncertainty, as described in Section~\ref{sec:systematics}, are included.
The expected contribution of \wzjjew events in the \wzjj signal region is about $13\%$ while $67\%$ of the events are expected to arise from \wzjjqcd production.
The expected contributions of polarised \wzjjew events are of the order of 1\%  for \wzjjewzz events and of 3.6\% for each of \wzjjewxz or \wzjjewzx events.

\begin{table}[t] %
\caption{
Observed and expected numbers of events in the \wzjjew signal region and in the two control regions, before the fit.
The expected number of polarised \wzjjewzz, \wzjjewzt, \wzjjewtz and \wzjjewtt events and the estimated number of background events from the other processes are shown.
The sum of the backgrounds containing misidentified leptons is labelled ``Misid. leptons''.
Numbers are provided separately for Run 2 and Run 3 data samples.
The total systematic uncertainties are reported.
\label{tab:EventYields_prefit}
}
\centering
\resizebox{0.98\textwidth}{!}{

\begin{tabular}{l *{6}{S[table-format=4.1]@{\,$\pm$\,}S[table-format=3.1]}}
\toprule
& \multicolumn{4}{c}{SR} & \multicolumn{4}{c}{$ZZ$-CR} & \multicolumn{4}{c}{$b$-CR} \\
\cmidrule(lr){2-5} \cmidrule(lr){6-9} \cmidrule(lr){10-13}
& \multicolumn{2}{c}{Run 2} & \multicolumn{2}{c}{Run 3} & \multicolumn{2}{c}{Run 2} & \multicolumn{2}{c}{Run 3} & \multicolumn{2}{c}{Run 2} & \multicolumn{2}{c}{Run 3} \\
\midrule
Data & \multicolumn{2}{c}{646} & \multicolumn{2}{c}{953} & \multicolumn{2}{c}{210} & \multicolumn{2}{c}{235} & \multicolumn{2}{c}{666} & \multicolumn{2}{c}{766} \\
Total pred. & 800 & 190 & 1060 & 270 & 204 & 11 & 240 & 24 & 687 & 56 & 751 & 75 \\
\midrule
$\wzjjewzz$ & 8.8 & 0.2 & 11.2 & 0.6 & \multicolumn{2}{c}{---} & \multicolumn{2}{c}{---} & \multicolumn{2}{c}{---} & \multicolumn{2}{c}{---} \\
$\wzjjewzt$ & 19.6 & 0.5 & 25.6 & 1.4 & \multicolumn{2}{c}{---} & \multicolumn{2}{c}{---} & \multicolumn{2}{c}{---} & \multicolumn{2}{c}{---} \\
$\wzjjewtz$ & 20.3 & 0.5 & 26.4 & 1.4 & \multicolumn{2}{c}{---} & \multicolumn{2}{c}{---} & \multicolumn{2}{c}{---} & \multicolumn{2}{c}{---} \\
$\wzjjewtt$ & 58.4 & 1.8 & 76.5 & 4.5 & \multicolumn{2}{c}{---} & \multicolumn{2}{c}{---} & \multicolumn{2}{c}{---} & \multicolumn{2}{c}{---} \\
\midrule
$\wzjjew$ & \multicolumn{2}{c}{---} & \multicolumn{2}{c}{---} & 0.6 & 0.1 & 0.8 & 0.2 & 5.1 & 1.3 & 3.4 & 0.9 \\
$\wzjjqcd$ & 530 & 190 & 710 & 260 & 6.3 & 2.1 & 8.7 & 3.0 & 93 & 36 & 75 & 27 \\
Misid. leptons & 46 & 11 & 62 & 19 & 1.7 & 0.4 & 2.2 & 0.5 & 155 & 38 & 195 & 61 \\
$\zzjjqcd$ & 41.6 & 2.9 & 51.8 & 9.4 & 158.4 & 8.6 & 181 & 20 & 10.1 & 0.6 & 9.0 & 0.9 \\
$\tzj$ & 28.5 & 1.0 & 37.3 & 3.5 & 0.4 & {\text{$<$0.1}} & 0.8 & 0.1 & 134.1 & 5.0 & 154.4 & 9.4 \\
$\ttV$ & 20.9 & 0.7 & 22.9 & 2.3 & 10.0 & 0.3 & 11.9 & 1.0 & 289 & 11 & 312 & 19 \\
$\zzjjew$ & 6.7 & 1.7 & 15.4 & 3.9 & 22.8 & 5.7 & 30.4 & 7.7 & 0.3 & 0.1 & 0.3 & 0.1 \\
$\wzjjint$ & 6.5 & 1.3 & 9.2 & 1.9 & 0.1 & {\text{$<$0.1}} & 0.1 & {\text{$<$0.1}} & 0.6 & 0.1 & 0.4 & 0.1 \\
$\wzjjew$ in $\tau$ & 5.9 & 0.6 & 7.5 & 0.8 & \multicolumn{2}{c}{---} & \multicolumn{2}{c}{---} & \multicolumn{2}{c}{---} & \multicolumn{2}{c}{---} \\
$\vvv$ & 2.4 & 0.6 & 3.7 & 0.9 & 4.1 & 1.0 & 4.8 & 1.2 & 0.4 & 0.1 & 0.4 & 0.1 \\
\bottomrule
\end{tabular}%


}
\end{table}


%
\section{Measurements methodology}
\label{sec:analysis}
%

\begin{table}[t]
\centering
\caption{Input variables used for the training of the different DNN classifiers. For each variable,  a mark $\checkmark$ means that the variable is used for a specific DNN.}
\label{tab:DNNVariable}
\resizebox{\textwidth}{!}{%
\begin{tabular}{llcccc}
\toprule
\textbf{Symbol} & \textbf{Description}
& \SvsBDNN & \PolZZDNN & \PolZXDNN & \PolXZDNN \\
\midrule
\multicolumn{6}{l}{\textit{Event topology}} \\
$\zeta_{\mathrm{lep.}}$ & Centrality of the $WZ$ system relative to the tagging jets & $\checkmark$ &   &   &   \\
\ayzlw & Rapidity difference between $Z$ boson and lepton from $W$ decay & $\checkmark$ & $\checkmark$ & $\checkmark$ & $\checkmark$ \\
$z_\ell$ & Zeppenfeld variable~\cite{Rainwater:1996ud} for leptons &   &   &   & $\checkmark$ \\
$N_{\mathrm{jets}}$ & Multiplicity of jets with $\pt > 25$~\GeV & $\checkmark$ &   &   &   \\
$R_{p_\mathrm{T}}^{\mathrm{hard}}$ & Ratio of vectorial to scalar $p_\mathrm{T}$ sum of jets & $\checkmark$ &   &   &   \\
\midrule
\multicolumn{6}{l}{\textit{Tagging jets}} \\
$p_\mathrm{T}^{j_1}$ & Transverse momentum of leading tagging jet     & $\checkmark$ &   &   & $\checkmark$ \\
$p_\mathrm{T}^{j_2}$ & Transverse momentum of sub-leading tagging jet &   &   &   & $\checkmark$ \\
$\eta^{j_1}$ & Pseudorapidity of leading tagging jet     &   & $\checkmark$ & $\checkmark$ & $\checkmark$ \\
$\eta^{j_2}$ & Pseudorapidity of sub-leading tagging jet & $\checkmark$ & $\checkmark$ & $\checkmark$ &   \\
$\phi^{j_2}$ & Azimuthal angle of sub-leading tagging jet &   &   & $\checkmark$ &   \\
$E^{j_1}$ & Energy of leading tagging jet & $\checkmark$ &   &   &   \\
$m_{jj}$ & Invariant mass of the dijet system  & $\checkmark$ &   &   &   \\
$\Delta\eta_{jj}$ & Pseudorapidity difference between the two tagging jets  & $\checkmark$ & $\checkmark$ & $\checkmark$ & $\checkmark$ \\
$\Delta\phi_{jj}$ & Azimuthal angle difference between the two tagging jets & $\checkmark$ & $\checkmark$ & $\checkmark$ &   \\
$\Delta R(j_1, Z)$ & Angular separation between the leading jet and the $Z$ boson &   & $\checkmark$ & $\checkmark$ & $\checkmark$ \\
$\langle p_\mathrm{T}^{jj} \rangle$ & Average transverse momentum of the two tagging jets &   &   &   & $\checkmark$ \\
$A_{p_\mathrm{T}^{jj}}$ & Asymmetry in $p_\mathrm{T}$ between the two tagging jets & $\checkmark$ & $\checkmark$ & $\checkmark$ &   \\
\midrule
\multicolumn{6}{l}{\textit{Boson kinematics}} \\
\ptw  & Transverse momentum of the $W$ boson    & $\checkmark$ & $\checkmark$ & $\checkmark$ & $\checkmark$ \\
\ptz  & Transverse momentum of the $Z$ boson    &   & $\checkmark$ & $\checkmark$ & $\checkmark$ \\
\ptwz & Transverse momentum of the $WZ$ system  & $\checkmark$ & $\checkmark$ & $\checkmark$ & $\checkmark$ \\
\mtwz & Transverse mass of the $WZ$ system  &   & $\checkmark$ & $\checkmark$ & $\checkmark$ \\
\rvtwoone & Ratio of leading to sub-leading boson $p_\mathrm{T}$ &   & $\checkmark$ &   &   \\
\midrule
\multicolumn{6}{l}{\textit{Lepton kinematics}} \\
$p_\mathrm{T}^{\ell_1^Z}$ & Transverse momentum of leading $Z$ lepton &   & $\checkmark$ &   & $\checkmark$ \\
$p_\mathrm{T}^{\ell_2^Z}$ & Transverse momentum of second $Z$ lepton  &   &   &   & $\checkmark$ \\
$p_\mathrm{T}^{\ell^W}$ & Transverse momentum of $W$ lepton   &   &   & $\checkmark$ & $\checkmark$ \\
$\eta^{\ell_1^Z}$ & Pseudorapidity of leading $Z$ lepton & $\checkmark$ & $\checkmark$ & $\checkmark$ & $\checkmark$ \\
$\eta^{\ell_2^Z}$ & Pseudorapidity of second $Z$ lepton  &   & $\checkmark$ &   & $\checkmark$ \\
$\eta^{\ell^W}$ & Pseudorapidity of $W$ lepton   & $\checkmark$ & $\checkmark$ & $\checkmark$ & $\checkmark$ \\
$E_\mathrm{T}^\mathrm{miss}$ & Missing transverse momentum &   &   & $\checkmark$ & $\checkmark$ \\
\midrule
\multicolumn{6}{l}{\textit{Angular and polarisation-sensitive observables}} \\
\ctw & Cosine of decay angle in the $W$ rest frame relative to the $WZ$ rest frame  &   & $\checkmark$ & $\checkmark$ &   \\
\ctz & Cosine of decay angle in the $Z$ rest frame relative to the $WZ$ rest frame &   & $\checkmark$ &   & $\checkmark$ \\
\ctv & Cosine of decay angle in the $WZ$ rest frame & $\checkmark$ & $\checkmark$ & $\checkmark$ &   \\
$\Delta\phi(\ell_{1}^{Z},\ell_{2}^{Z})$ & Azimuthal angle difference between leptons from the $Z$ & $\checkmark$ & $\checkmark$ &   &   \\
$\Delta\phi(\ell^{W},\nu)$ & Azimuthal angle difference between lepton and neutrino from the $W$ & $\checkmark$ &   & $\checkmark$ &   \\
\bottomrule
\end{tabular}
}
\end{table}

Given the small contribution to the signal region of 13\% from \wzjjew processes, an efficient separation of the \wzjjew contribution from \wzjjqcd and other background events is required.
A multivariate discriminant based on a deep neural network (DNN) classifer, labelled ``\SvsBDNN'' DNN,  is trained to separate \wzjjew events from other background SM processes.
A set of 18 kinematical observables is used, related to the event topology, the kinematics of the leptons, tagging jets and vector bosons, and the specific angular observables sensitive to polarisation, as detailed in Table~\ref{tab:DNNVariable}.
The variables related to leptons are the pseudorapidities of the leading lepton for the $Z$ boson decay, $\eta^{\ell_1^Z}$, and of the lepton from the $W$ boson decay, $\eta^{\ell^W}$; the absolute value of the cosine of the scattering angle of the $W$ or $Z$  boson in the $WZ$ centre-of-mass frame, \ctv, calculated with respect to the beam axis~\cite{Baur:1994ia,Franceschini:2017xkh}; the difference in azimuthal angle between the two leptons from the $Z$ boson, $\Delta\phi(\ell_{1}^{Z},\ell_{2}^{Z})$, and between the lepton and the neutrino from the $W$ boson, $\Delta\phi(\ell^{W},\nu)$.
The variables related to the kinematic properties of the two tagging jets are the invariant mass of the two jets, $m_{jj}$, the transverse momentum, \ptji, and energy, $E^{j_1}$,  of the leading jet,  the difference in pseudorapidity and azimuthal angle between the two jets, $\Delta \eta_{jj}$ and $\Delta \phi_{jj}$, the pseudorapidity of the sub-leading jet, the jet multiplicity and the asymmetry in $p_\mathrm{T}$ between the two tagging jets $A_{p_\mathrm{T}^{jj}} = (\ptji - \ptjii ) / ( \ptji + \ptjii )$.
The variables related to the kinematic properties of the vector bosons are the transverse momenta of the $W$ and $Z$ bosons, the absolute difference between the rapidities of the $Z$ boson and the lepton from the decay of the $W$ boson, \ayzlw, and the transverse momentum of the \wz system \ptwz.
The variables that relate the kinematic properties of jets and leptons are the event balance $R_{p_\mathrm{T}}^{\mathrm{hard}}$, defined as the transverse component of the vector sum of the $WZ$ bosons and tagging jets momenta, normalised to their scalar \pt sum, and the centrality of the $WZ$ system relative to the tagging jets, defined as $\zeta_{\mathrm{lep.}} = \text{min}(\Delta\eta_{-},\Delta\eta_{+})$, with $\Delta\eta_{-} = \text{min}(\eta_{\ell}^{W},\eta_{\ell_2}^{Z},\eta_{\ell_1}^{Z}) - \text{min}(\eta_{j_1},\eta_{j_2})$ and $\Delta\eta_{+} = \text{max}(\eta_{j_1},\eta_{j_2}) - \text{max}(\eta_{\ell}^{W},\eta_{\ell_2}^{Z},\eta_{\ell_1}^{Z})$, where $\eta_{\ell_1}^{Z}$ and $\eta_{\ell_2}^{Z}$ are the pseudorapidity of the leading and sub-leading leptons associated to the $Z$ boson.
The good modelling by MC simulations of the distribution shapes and of the correlations of all input variables to the DNN is verified in a validation region defined by requiring $150 < \mjj < 500$~\GeV.
Distributions of the most important variables were also measured in the signal region and compared to MC predictions using the 13~\TeV data sample~\cite{ATLAS:2024ini}.

\begin{figure}[!htbp]
\centering

\subfloat[]{
\includegraphics[width=0.45\textwidth]{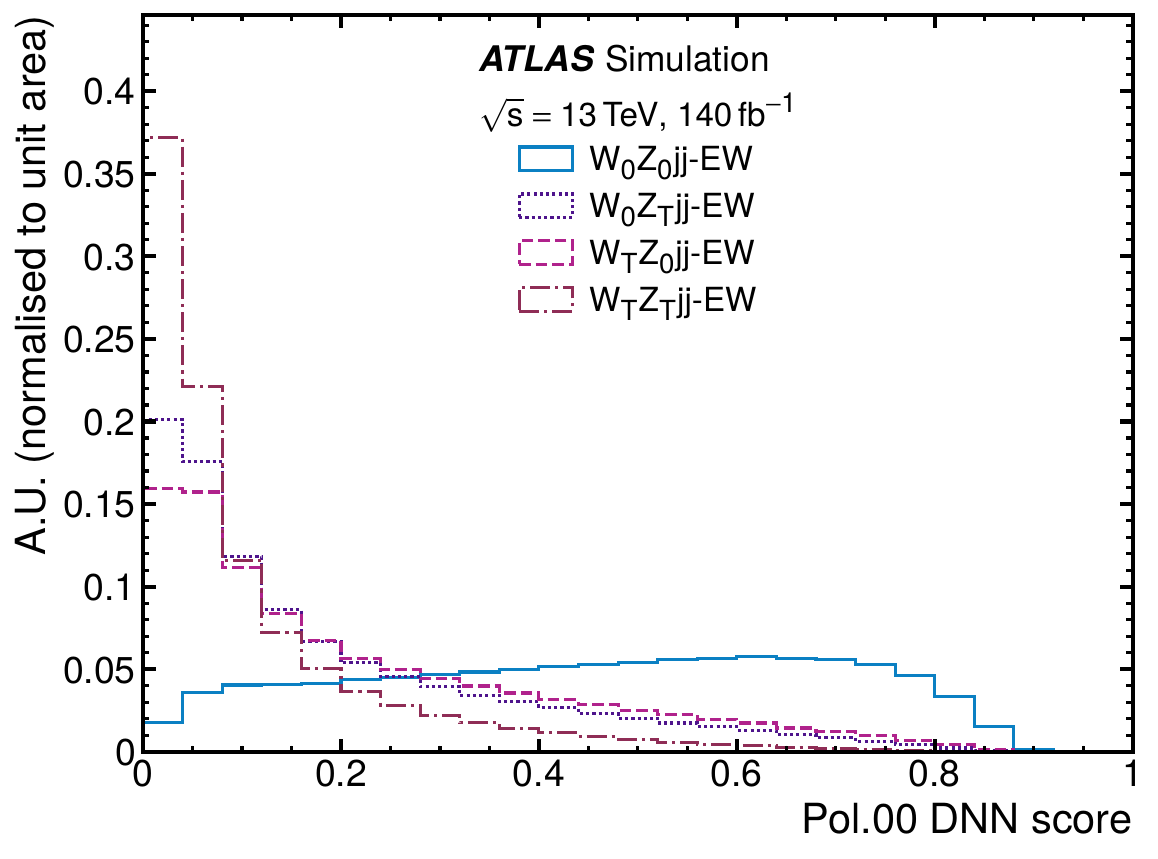}
}
\subfloat[]{
\includegraphics[width=0.45\textwidth]{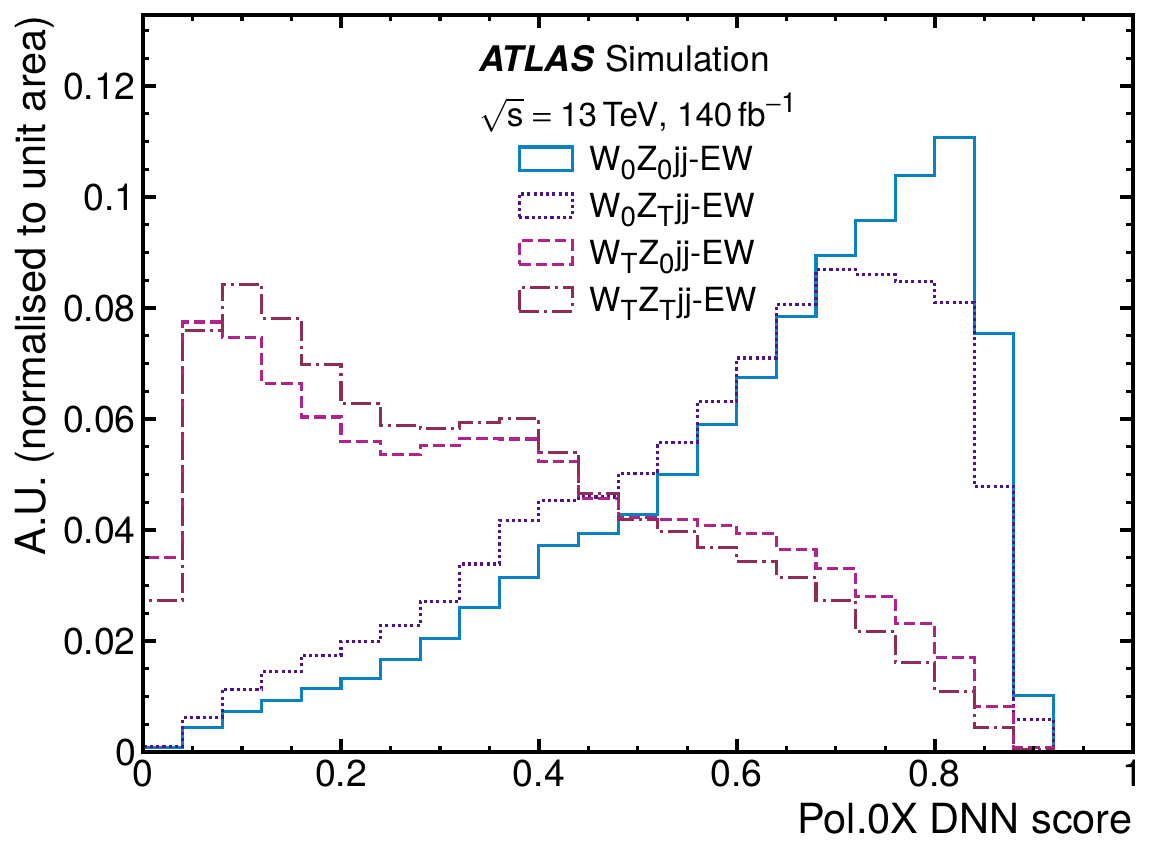}
}\\
\subfloat[]{
\includegraphics[width=0.45\textwidth]{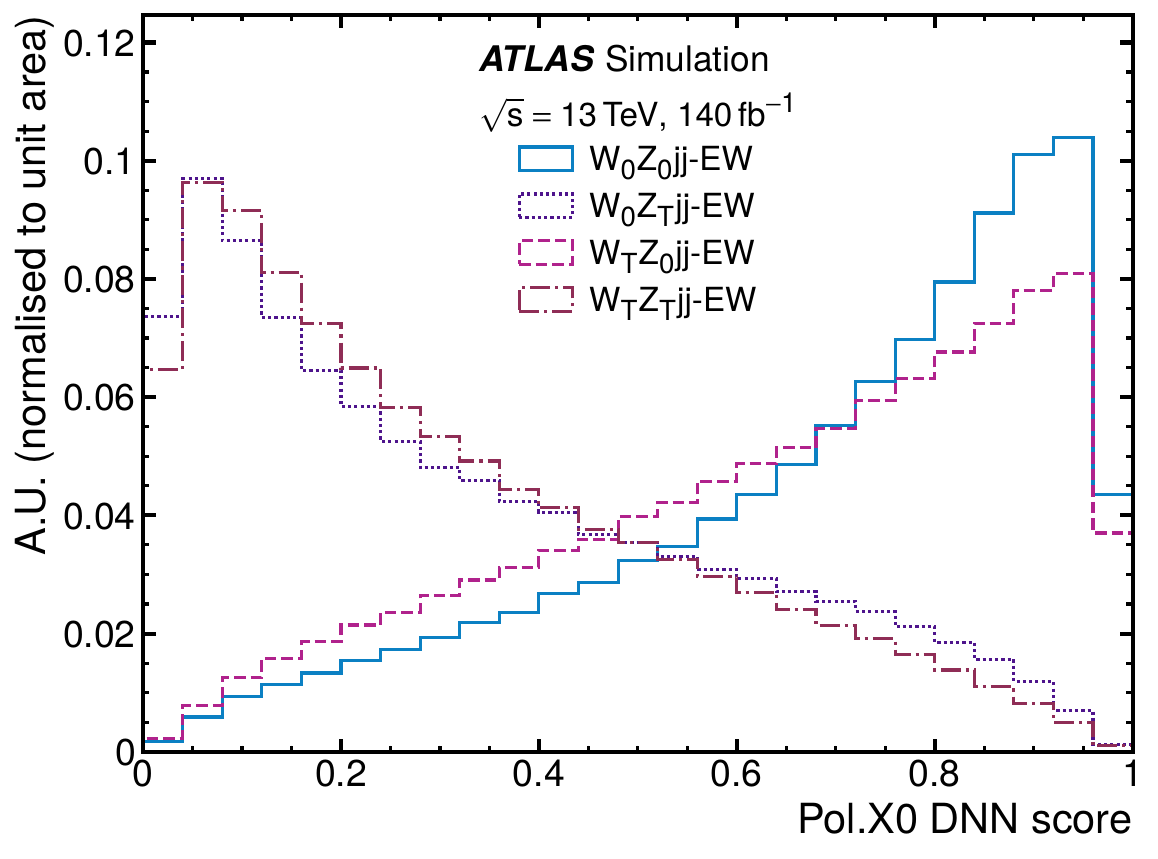}
}

\caption{Example distributions of the scores of the (a) \PolZZDNN (b) \PolZXDNN and (c) \PolXZDNN  DNNs for \wzjjewzz, \wzjjewzt, \wzjjewtz and \wzjjewtt events from the \Sherpa 3.0 MC.
All distributions are normalised to the same integral.
}
\label{fig:PolDNNScores}
\end{figure}

Among \wzjjew events, the expected fraction of \wzjjewxz, \wzjjewzx and \wzjjewzz events are  27\%, 28\% and 8\%,  respectively.
To reach an efficient separation of each of these three contributions from all \wzjjew events, three different multivariate discriminants based on DNN classifiers are used.
Each DNN is trained to separate each of the X0, 0X and 00 polarisation states, respectively, from all other polarisation states.
The three DNNs are labelled  ``\PolXZDNN'', ``\PolZXDNN'' and ``\PolZZDNN'', respectively.
The variables used as inputs of these three polarisation DNNs are detailed in Table~\ref{tab:DNNVariable}.
Among the variables used, \ctv brings sensitivity to joint longitudinal polarisation states.
The angular distributions of the decay products~\cite{WZPol36fb,ATLAS:2022oge},  \ctw and \ctz, of the $W$ and $Z$ gauge bosons, respectively, are sensitive to individual polarisation of the bosons.
The observable $\theta^*_{\ell^W}$ is defined as the decay angle of the charged lepton in the $W$ rest frame relative to the $W$ direction in the \wz centre-of-mass frame.
The pseudorapidity of the $W$ boson is reconstructed using an estimate of the longitudinal momentum of the neutrino obtained using the neural-network based method developed in Ref.~\cite{ATLAS:2022oge}.
The observable $\theta^*_{\ell^Z}$ is defined as the decay angle of the negatively charged lepton in the $Z$ rest frame relative to the $Z$ direction in the \wz centre-of-mass frame.
In VBS events, the kinematic of the two tagging jets, and in particular their average transverse momentum, $\langle p_\mathrm{T}^{jj} \rangle =  (\ptji + \ptjii )/2$, and $A_{p_\mathrm{T}^{jj}}$ are also sensitive to the joint polarisation of the scattering vector bosons~\cite{Brehmer:2014pka,Denner:2025xdz}.
Input variables of each DNN are selected based on a larger set of discriminating observables and retaining only those improving the signal-to-background separation.
The DNN classifiers achieve integrals of the receiver operating characteristic (ROC) curves of 0.84, 0.80 and 0.84 for separating the X0 from XT, 0X from TX, and 00 from T0+0T+TT, states, respectively.
Example distributions of the output score of each DNN classifier for polarised \wzjjew MC events  are presented in Figure~\ref{fig:PolDNNScores}.

For a simultaneous separation of \wzjjew events from other SM processes and of the polarisation state of interest  from other polarisation states of \wzjjew events, a two-dimensional plane is formed from the score of the \SvsBDNN and \PolXZDNN (or \PolZXDNN, \PolZZDNN) DNN classifiers.
In this plane, events of similar $S/\sqrt{B}$ ratio are grouped in ten clusters, where $S$ and $B$ are the numbers of signal and background events, respectively, with the signal corresponding to targeted polarised events.
The number of clusters results from a compromise between having enough granularity to exploit the discrimination power of the DNNs and avoiding too large statistical uncertainty on the expected number of signal and background events in each cluster.
The \textsc{KMeans} clustering algorithm~\cite{kmeans,kmeans2} from the \texttt{scikit-learn} package~\cite{Pedregosa:2011ork} is used as a first step of this grouping.
Voronoi~\cite{Voronoi} regions naturally arise from the resulting clusters.
An example of the Voronoi regions used for the measurement of \fXz is shown in Figure ~\ref{fig:VoronoiRegions}.
The content of the ten Voronoi regions is then represented in a one-dimensional histogram where the regions are numbered according to their increasing fraction of signal events.
This representation is referred to as ``\fitDiscriName'' and is shown in Figures~\ref{fig:ScoreDistributions_Run2} and~\ref{fig:ScoreDistributions_Run3}.

\begin{figure}[!htbp]
\centering
\includegraphics[width=0.45\textwidth]{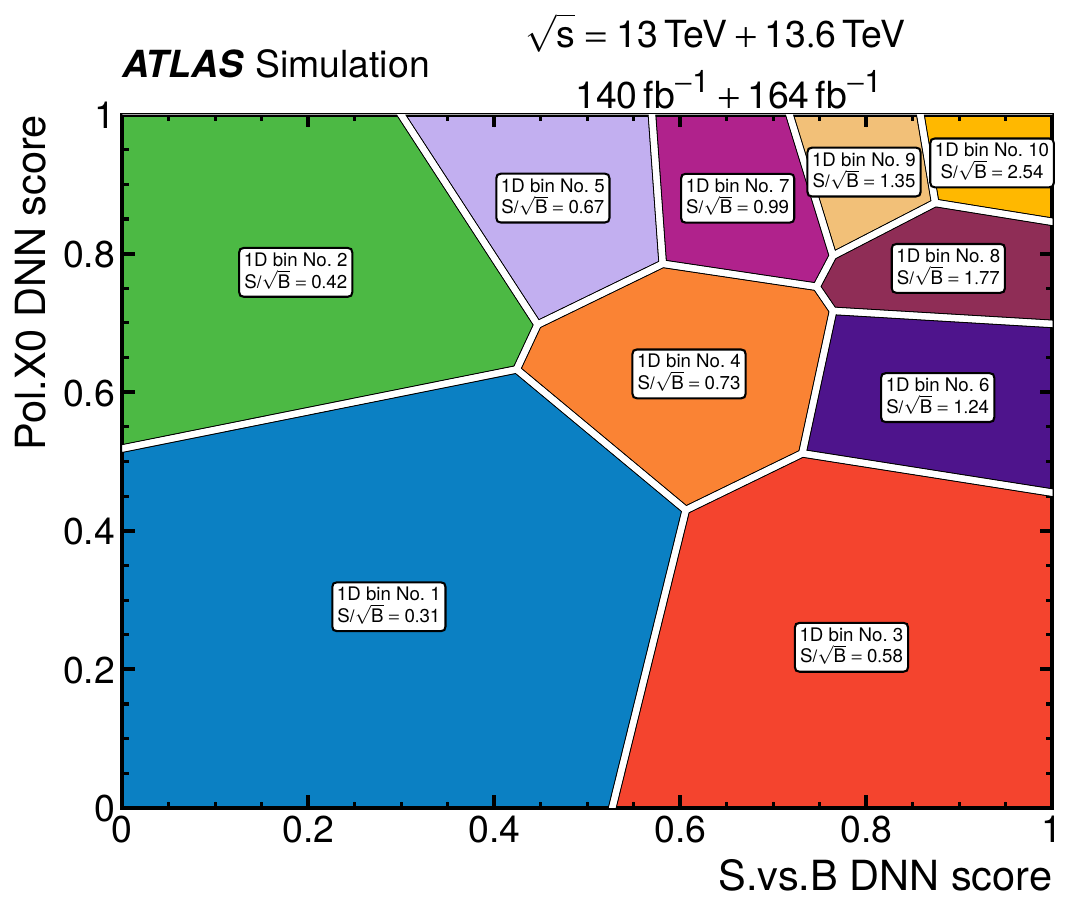}

\caption{Voronoi regions used for the \fXz measurement. The global $S/\sqrt{B}$ ratio of each region is indicated. The bin number in the \fitDiscriName corresponding to each region is indicated.
}
\label{fig:VoronoiRegions}
\end{figure}

To better separate the \ttV and \tzj events in the \bCR and to constrain the normalisation of both contributions using data in this CR, a boosted decision tree (BDT) discriminant, labelled \CRbBDT, is used, as defined in Ref.~\cite{ATLAS:2024ini}.
In the \ZZCR the invariant mass of the two tagging jets constraints the normalisation of the \zzjjqcd background and, to a lesser extent, the contribution of \zzjjew events.

To measure polarisation fractions, a binned maximum-likelihood fit to data in the SR  of the four \wzjjewzz, \wzjjewzt, \wzjjewtz and \wzjjewtt polarisation states templates is performed.
The \textsc{TRExFitter} framework~\cite{TRExFitter} is used.
An extended likelihood is built from the product of three likelihoods corresponding to the \fitDiscriName distribution in the SR, the \CRbBDT distribution in the \bCR and the $m_{jj}$ distribution in the \ZZCR.
The normalisations of these backgrounds are introduced in the likelihood as parameters, labelled  \muttV, \mutZj and \muZZ for \ttV, \tzj and \zzjjqcd backgrounds, respectively.
The shapes of these backgrounds are taken from MC predictions and can vary within the uncertainties affecting the measurement as described in Section~\ref{sec:systematics}.
The inclusion of the two control regions in the fit allows the yields of the \ttV, \tzj and \zzjjqcd backgrounds to be constrained by data.
They are treated as unconstrained nuisance parameters that are determined mainly by the data in the respective control region.
The normalisation and shape of the other irreducible backgrounds are taken from MC simulations.
In particular, owing to the good separation between the \wzjjqcd and \wzjjew processes provided by the \SvsBDNN DNN, the normalisation of the \wzjjqcd background can be constrained by the fit using data of the SR.
The distribution of the reducible background is estimated from data using the matrix method presented in Section~\ref{sec:background}.
All background templates are allowed to vary in shape and normalisation within their respective uncertainties.
Each source of systematic uncertainty is implemented in the likelihood function as a nuisance parameter with a Gaussian constraint.

For the separate measurements of each polarisation fraction  \fXz, \fzX and \fzz, three independent fits are performed using their dedicated Voronoi regions.
The two independent parameters of each fit are the targeted particle-level fraction and the integrated fiducial cross-section of \wzjjew production, \sigew.
In each fit, the sum of the measured and non-measured polarisation fractions is enforced to be one.
By definition, each polarisation fraction is constrained to be between 0 and 1.
For each polarisation fraction measurement, the respective proportions of polarisation states among the non-measured ones  are taken from the MC prediction.
Fits to data allowing the respective proportions of non-measured polarisation states to vary were also tested and yield consistent results.
Fits to pseudo-data with injected variations of the non-measured polarisation fractions, deviating from the SM expectation, were performed.
In all cases, the fits are able to recover the true measured polarisation fractions.
This robustness of the fits is explained in the \fzX and \fXz cases by the very similar shapes of the DNN score distributions for the non-measured polarisation states, as observed in Figures~\ref{fig:PolDNNScores}(b) and (c).
In the \fzz case, the very good separation of the 00 state from the other polarisation states provided by the DNN, as shown in  Figures~\ref{fig:PolDNNScores}(a),  explains it.
From \MoCaNLO~\cite{Denner:2026phn,Denner:2025xdz} and \Sherpa 3.0 predictions, polarisation fractions are found to be the same for $pp$ collisions at $\sqrt{s} = 13$~\TeV\ and  $\sqrt{s} = 13.6$~\TeV within the statistical uncertainties of the calculations, which are less than 1\%.
Data recorded at $\sqrt{s} = 13$~\TeV\ and  $\sqrt{s} = 13.6$~\TeV\ are therefore combined in a simultaneous profile-likelihood fit as coming from two separate measurements of the same polarisation fractions.
The integrated fiducial cross-section is replaced by the ratio of the measured to the predicted cross-sections, $\muew = \sigew / \sigew^{\mathrm{Sherpa}}$, where $\sigew^{\mathrm{Sherpa}}$ is the cross-section predicted by \Sherpa 3.0.
This allows the combination of the two data samples at $\sqrt{s} = 13$~\TeV\ and  $\sqrt{s} = 13.6$~\TeV\ using the same parameter \muew.
As the $Z$ boson is easier to reconstruct than the $W$ boson, the measurement of \fXz is expected to have the highest experimental sensitivity among the three targeted polarisation fractions.


%
\clearpage
\section{Systematic uncertainties}
\label{sec:systematics}
%

Systematic uncertainties are considered in the signal and control regions that affect the shape and normalisation of the \fitDiscriName, the \mjj and \CRbBDT distributions for the individual backgrounds, and the acceptance of the polarised \wzjjew events and the shape of their templates.
Systematic uncertainties in the theoretical modelling by the event generator used to evaluate the \wzjjqcd, \wzjjint and polarised \wzjjew templates are considered.
Uncertainties due to higher-order QCD corrections are evaluated by varying the renormalisation and factorisation scales independently by factors of two and one-half, removing combinations where the variations differ by a factor of four.
These are referred to as QCD scale uncertainties.
It was shown in Ref.~\cite{ATLAS:2024ini} that the QCD scale uncertainties defined using this prescription are covering possible differences between data and the \Sherpa MC prediction in the DNN input variables, in particular for \wzjjqcd background events.
In the SR, these uncertainties are of up to $\pm 40\%$ and $\pm 10\%$ on the normalisation of the \fitDiscriName  distribution for \wzjjqcd and \wzjjint  events, respectively, and of up to $\pm 4\%$  on the shape of the \fitDiscriName  distribution for \wzjjew events.
The uncertainties in \wzjjqcd, \wzjjint and \wzjjew predictions due to the PDF and the $\alphas$ value used in the PDF determination are evaluated using the PDF4LHC prescription~\cite{Butterworth:2015oua}.
They are of the order of $1\%$ in shape for \wzjjew events.
A global modelling uncertainty in the \wzjjqcd background template that includes effects of the parton shower, hadronisation and underlying event models is estimated by comparing predictions of the \fitDiscriName distribution in the signal region from the \MGPy8 and \sherpa MC event generators.
The difference between the predicted shapes of the \fitDiscriName distribution from the two generators is considered as an uncertainty and affects the shape and normalisation of the \fitDiscriName distribution of \wzjjqcd events by up to $\pm 30\%$ at larger values of the \fitDiscriName distribution.
Similarly, a global modelling uncertainty in the polarised \wzjjew templates is estimated by comparing predictions of the \fitDiscriName distribution in the signal region from MC simulations from the \MGPy8 and \sherpa MC event generators.
This global modelling uncertainty includes effects of the parton shower, hadronisation and underlying event models, as well as differences in the polarised \wzjjew templates between \LO and 0,1j@LO simulations.
This global modelling uncertainty affects the shape of the \fitDiscriName distributions of polarised \wzjjew events by at most $20\%$ in regions corresponding to lower values of the \SvsBDNN DNN score.
Each of the uncertainties from higher-order QCD corrections and global modelling effects is independent and  is considered as correlated among the polarised \wzjjew templates
All these systematic uncertainties in theoretical modelling are considered as correlated between Run~2 and Run~3 samples.

The effect of \NLO EW corrections on the shape of polarised \wzjjewzz, \wzjjewzt, \wzjjewtz and \wzjjewtt  templates is studied using \MoCaNLO calculations~\cite{Denner:2025xdz,Denner:2026phn}.
The prediction from \MoCaNLO at \LO and \NLO EW accuracy is determined specifically for the \fitDiscriName distribution by implementing the evaluation of the DNN scores and the Voronoi regions in the \MoCaNLO code.
The \NLO EW corrections are found to have no effect at parton-level on the shape of the output distribution of the polarisation DNNs \PolXZDNN, \PolZXDNN and \PolZZDNN but only affect the shape of the \SvsBDNN background DNN by at most $+15$\% and $-28$\% at low and high values of the DNN score, respectively.
The parton-level \NLO EW corrections from \MoCaNLO in the \fitDiscriName are transformed to detector-level by reweighting \Sherpa 3.0 MC events at particle level.
The effect of \NLO EW corrections on the shape of the \fitDiscriName detector-level distributions is found to be of maximum size of 20\% for all four polarisation states.
The difference between the predicted shapes of the \fitDiscriName with and without \NLO EW corrections is considered as an uncertainty affecting the shape of each \wzjjewzz, \wzjjewzt, \wzjjewtz or \wzjjewtt  templates.
This uncertainty is considered as correlated between the polarised templates and between Run~2 and Run~3 samples.

Systematic uncertainties affecting the reconstruction, identification and energy calibration of jets, electrons and muons are propagated through the analysis.
The dominant source of uncertainty is the jet energy calibration and the modelling of pile-up.
The uncertainties in the jet energy are obtained from $\sqrt{s} = 13$~\TeV\ simulations and {\textit{in situ}} measurements~\cite{JETM-2018-05}.
The uncertainties in the jet energy resolution~\cite{JETM-2018-05}, in the suppression of jets originating from pile-up~\cite{Aad:2015ina}, in the $b$-tagging efficiency and in the mistag rate~\cite{FTAG-2019-07} are also considered.
The uncertainty in the \MET measurement is estimated by propagating the uncertainties in the transverse momenta of hard physics objects and by applying momentum scale and resolution uncertainties to the track-based soft term~\cite{JETM-2020-03}. %
A variation in the pile-up reweighting of MC events is included to cover the uncertainty in the ratio of the predicted and measured $pp$ inelastic cross-sections~\cite{Aaboud:2016mmw}.
The uncertainties due to lepton reconstruction, identification, isolation requirements and trigger efficiencies as well as in the lepton momentum and energy scale and resolution are assessed using tag-and-probe methods in $Z \to \ell \ell$ events~\cite{ATLAS:2019qmc,ATLAS:2020auj}.
The detector-related uncertainties are considered as uncorrelated between Run 2 and Run 3 data samples, except for a few jet uncertainties which are fully correlated between Run 2 and Run 3 measurements.

The uncertainty in the amount of background from misidentified leptons takes into account the limited number of events in the control regions as well as the difference in background composition between the control region used to determine the lepton misidentification rate and the control regions used to estimate the yield in the signal region.
This results in an uncertainty of about $35\%$ in the total  misidentified-lepton background yield and in the shape of the differential distributions of the reducible background events.
Uncertainties due to the theoretical modelling of \zz generated events are considered.
They arise from higher-order QCD corrections and the PDFs and are evaluated in the same way as for \wz\ events.
The uncertainty due to irreducible background sources other than \zz, \ttV or \tzj, that are not constrained by data in the dedicated CRs, is evaluated by propagating the uncertainty in their MC cross-sections.
These are $25\%$ for $VVV$~\cite{ATL-PHYS-PUB-2016-002} %
and for \zzjjew to account for the potentially large impact of scale variations.
A normalisation uncertainty of $10$\% is attributed to the contribution of events from \wzjjew production with at least one $\tau$-lepton decay to account for a possible mis-modelling in the MC simulation of the efficiency to select \wzjjew events with a $\tau$-lepton decay.
It is estimated from differences between the \Sherpa and \Madgraph simulations.
The background-related uncertainties are considered as uncorrelated between Run 2 and Run 3 data samples.

The uncertainties in the integrated luminosities for data collected in Run 2 and Run 3 are $0.83\%$~\cite{DAPR-2021-01} and $1.9\%$~\cite{ATL-DAPR-PUB-2025-001}, respectively, obtained using the LUCID-2 detector~\cite{LUCID2} for the primary luminosity measurements, complemented by measurements using the inner detector and calorimeters.

The effect of the systematic uncertainties on the final results after the maximum-likelihood fit is shown in Table~\ref{tab:Results:SystematicSummary} where the breakdown of the contributions to the uncertainties in the measured  polarisation fractions and fiducial cross-section normalisation factor is presented.

\begin{table}[!htbp]
\caption{
Summary of the relative uncertainties in the measured values of the polarisation fraction \fXz and of \muew.
The uncertainties are reported as percentages.
For \muew, values from the fit of \fXz are used.
The contribution from the modelling of background events containing misidentified leptons is labelled ``Misid. leptons background''.
The contribution from the modelling of all irreducible background events except \wzjjqcd and \wzjjint processes is labelled ``Other backgrounds''.
}
\label{tab:Results:SystematicSummary}
\centering
%

\begin{tabular}{l S[table-format=2.2] S[table-format=2.2]}
\toprule
& {$f_{X0}$} & {$\muew$} \\
\midrule
\multicolumn{3}{c}{Relative uncertainty [\%]} \\
\midrule
\wzjjew Theory modelling & 1.7 & 4.9 \\
\wzjjqcd Theory modelling & 1.6 & 5.0 \\
\wzjjint Theory modelling & {\text{$<$0.1}} & 0.4 \\
PDFs & 0.3 & 0.7 \\
\midrule
Electrons & 1.3 & 1.6 \\
$b$-tagging & 0.5 & 0.5 \\
Jets & 3.6 & 6.6 \\
$E_{\mathrm{T}}^{\mathrm{miss}}$ & 0.3 & 0.5 \\
Muons & 0.8 & 0.9 \\
Pile-up & 0.5 & 1.9 \\
Other backgrounds & 0.8 & 1.3 \\
Misid. leptons background & 0.6 & 1.0 \\
MC statistics & 2.4 & 1.1 \\
\midrule
All systematics & 5.2 & 10.2 \\
Luminosity & 0.4 & 1.2 \\
Statistics & 26.6 & 13.8 \\
\midrule
Total & 27.1 & 17.2 \\
\bottomrule
\end{tabular}


\end{table}


%
\section{Results}
\label{sec:xsection_results}
%

\begin{table}[!htbp] %
\caption{Observed and expected numbers of events in the \wzjjew signal region and in the two control regions, after the fit for the \fXz measurement.
The expected number of polarised \wzjjewxz and \wzjjewxt events and the estimated number of background events from the other processes are shown.
The sum of the backgrounds containing misidentified leptons is labelled ``Misid. leptons''.
The total correlated post-fit systematic uncertainties are quoted.
Numbers are provided separately for Run 2 and Run 3 data samples.
\label{tab:EventYields_postfit}}
\centering
\resizebox{0.98\textwidth}{!}{

\begin{tabular}{l *{6}{S[table-format=4.1]@{\,$\pm$\,}S[table-format=3.1]}}
\toprule
& \multicolumn{4}{c}{SR} & \multicolumn{4}{c}{$ZZ$-CR} & \multicolumn{4}{c}{$b$-CR} \\
\cmidrule(lr){2-5} \cmidrule(lr){6-9} \cmidrule(lr){10-13}
& \multicolumn{2}{c}{Run 2} & \multicolumn{2}{c}{Run 3} & \multicolumn{2}{c}{Run 2} & \multicolumn{2}{c}{Run 3} & \multicolumn{2}{c}{Run 2} & \multicolumn{2}{c}{Run 3} \\
\midrule
Data & \multicolumn{2}{c}{646} & \multicolumn{2}{c}{953} & \multicolumn{2}{c}{210} & \multicolumn{2}{c}{235} & \multicolumn{2}{c}{666} & \multicolumn{2}{c}{766} \\
Total pred. & 668 & 23 & 934 & 28 & 210 & 14 & 238 & 15 & 663 & 26 & 771 & 28 \\
\midrule
$\wzjjewxz$ & 36 & 11 & 48 & 14 & \multicolumn{2}{c}{---} & \multicolumn{2}{c}{---} & \multicolumn{2}{c}{---} & \multicolumn{2}{c}{---} \\
$\wzjjewxt$ & 76 & 18 & 103 & 24 & \multicolumn{2}{c}{---} & \multicolumn{2}{c}{---} & \multicolumn{2}{c}{---} & \multicolumn{2}{c}{---} \\
\midrule
$\wzjjew$ & \multicolumn{2}{c}{---} & \multicolumn{2}{c}{---} & 0.6 & 0.1 & 0.8 & 0.2 & 5.0 & 1.3 & 3.4 & 0.9 \\
$\wzjjqcd$ & 393 & 30 & 573 & 37 & 5.0 & 0.4 & 6.8 & 0.6 & 62.2 & 10.0 & 59.1 & 5.0 \\
Misid. leptons & 45 & 10 & 54 & 15 & 1.7 & 0.4 & 2.2 & 0.5 & 152 & 36 & 168 & 48 \\
$\zzjjqcd$ & 44.0 & 4.0 & 52.4 & 5.4 & 170 & 15 & 183 & 16 & 9.9 & 0.6 & 9.3 & 0.5 \\
$\tzj$ & 33.3 & 6.4 & 44.1 & 8.9 & 0.5 & 0.1 & 1.0 & 0.2 & 157 & 30 & 182 & 35 \\
$\ttV$ & 19.8 & 2.4 & 25.7 & 3.6 & 9.5 & 1.2 & 13.3 & 1.8 & 275 & 32 & 349 & 42 \\
$\zzjjew$ & 5.6 & 1.3 & 13.5 & 3.2 & 19.1 & 4.6 & 26.7 & 6.3 & 0.2 & 0.1 & 0.2 & 0.1 \\
$\wzjjint$ & 6.5 & 1.3 & 9.4 & 1.9 & 0.1 & {\text{$<$0.1}} & 0.1 & {\text{$<$0.1}} & 0.6 & 0.1 & 0.4 & 0.1 \\
$\wzjjew$ in $\tau$ & 5.9 & 0.6 & 7.6 & 0.8 & \multicolumn{2}{c}{---} & \multicolumn{2}{c}{---} & \multicolumn{2}{c}{---} & \multicolumn{2}{c}{---} \\
$\vvv$ & 2.4 & 0.6 & 3.7 & 0.9 & 4.1 & 1.0 & 4.9 & 1.2 & 0.4 & 0.1 & 0.4 & 0.1 \\
\bottomrule
\end{tabular}%


}
\end{table}

The post-fit distributions of the  \fitDiscriName in the signal region used for the \fzz, \fzX and \fXz measurements are presented in Figures~\ref{fig:ScoreDistributions_Run2} and \ref{fig:ScoreDistributions_Run3} for the Run 2 and Run 3 data samples, respectively.
The background normalisations, signal normalisation and nuisance parameters are adjusted by the profile-likelihood fit.
The post-fit distributions of \mjj and of the \CRbBDT score in the \ZZCR and \bCR, respectively, are also presented in Figures~\ref{fig:ScoreDistributions_Run2} and \ref{fig:ScoreDistributions_Run3}.
For these distributions, the result from the \fXz fit is shown, as it is expected to provide the most sensitive measurement of a polarisation fraction.
The results of the \fzz and \fzX fits are similar.
The corresponding post-fit event yields are detailed in Table~\ref{tab:EventYields_postfit}.
The normalisation parameters of the \ttV, \tzj and \zz backgrounds constrained by data in the control and signal regions are measured to be  $\muttV = 0.96 \pm 0.12$ ($1.09 \pm 0.14$), $\mutZ = 1.18 \pm 0.22$ ($1.15 \pm 0.22$) and $\muZZ = 1.09 \pm 0.11$ ($0.97 \pm 0.11$) for Run 2 (Run 3) data. %
The \wzjjqcd background is scaled in the fit by factors of  $0.74 \pm 0.06$ and $0.81 \pm 0.05$ for Run 2 and Run 3 data, respectively.
These values are consistent with the normalisation uncertainty of $^{+40}_{-30}\%$ due to higher order QCD corrections affecting the \wzjjqcd MC sample.
These renormalisation factors are also consistent with previous measurements~\cite{ATLAS:2024ini,ATLAS:2023sua}.
No uncertainties other than the \wzjjqcd QCD scale uncertainty are significantly constrained or pulled in the fit.

\begin{figure}[t]
\centering

\subfloat[]{
\includegraphics[width=0.33\textwidth]{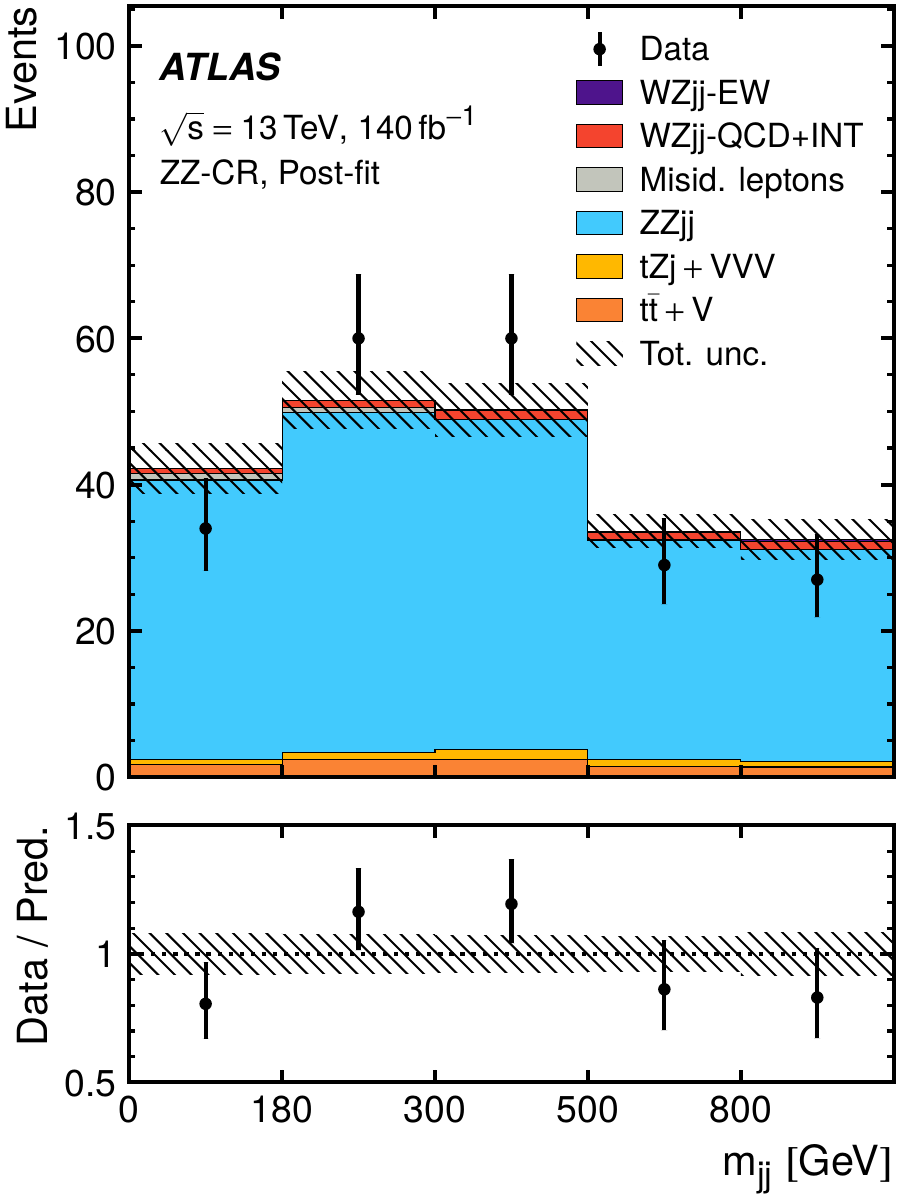}
}
\subfloat[]{
\includegraphics[width=0.33\textwidth]{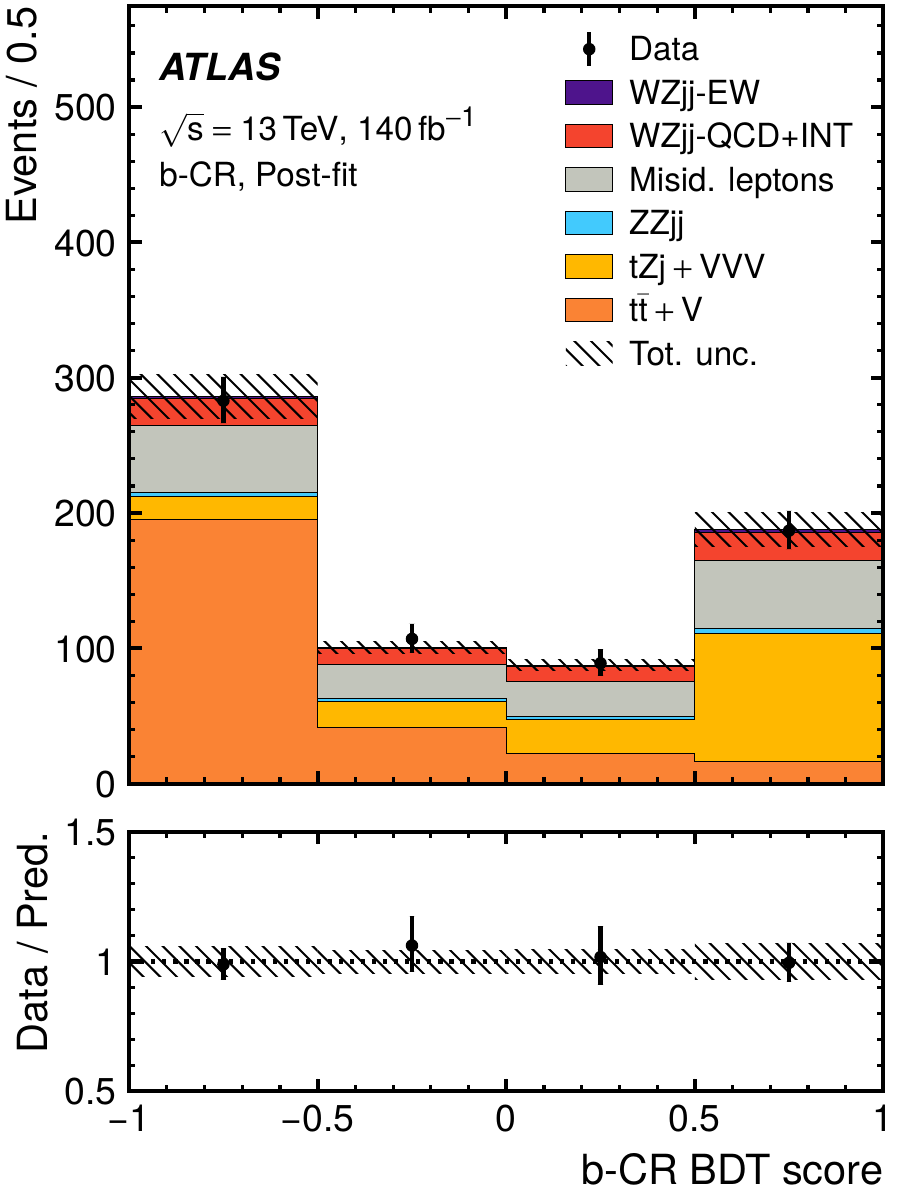}
}\\

\subfloat[]{
\includegraphics[width=0.33\textwidth]{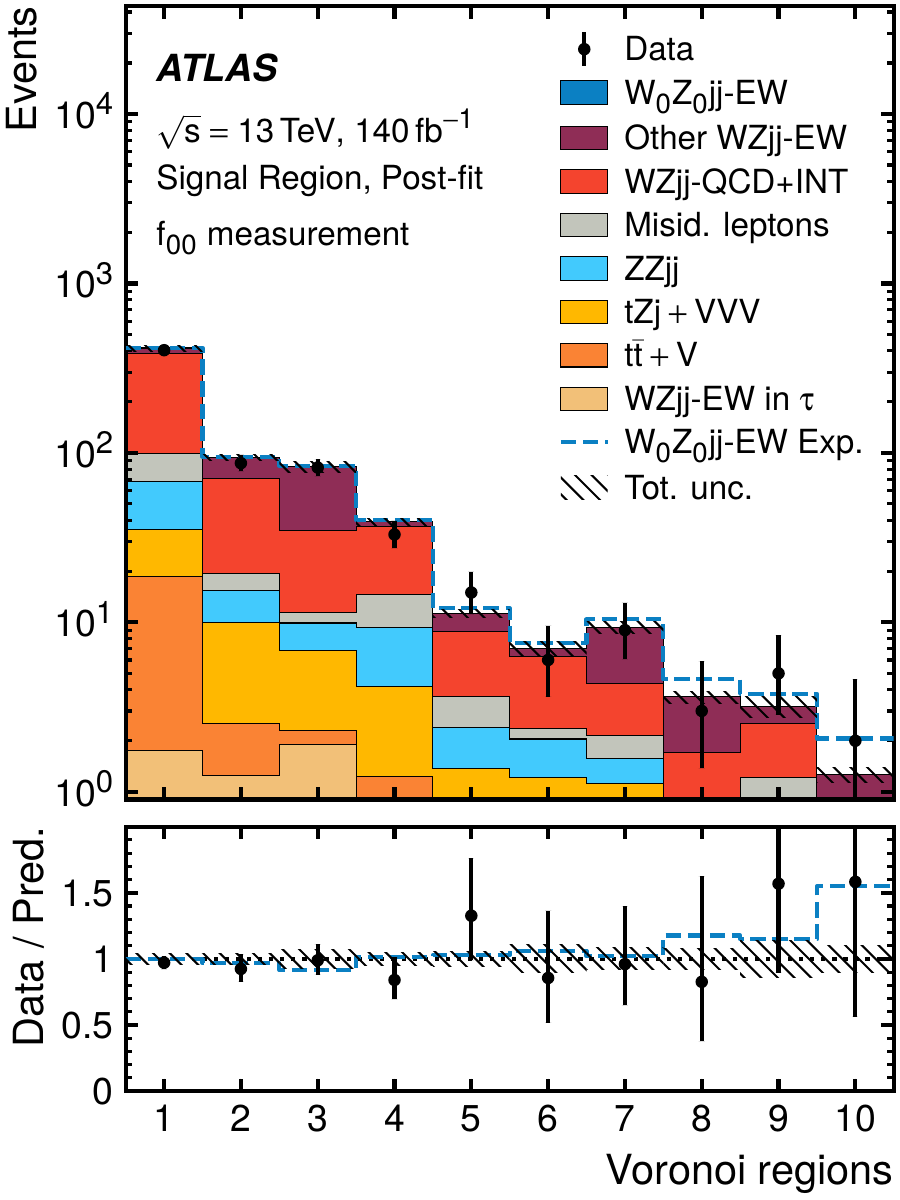}
}
\subfloat[]{
\includegraphics[width=0.33\textwidth]{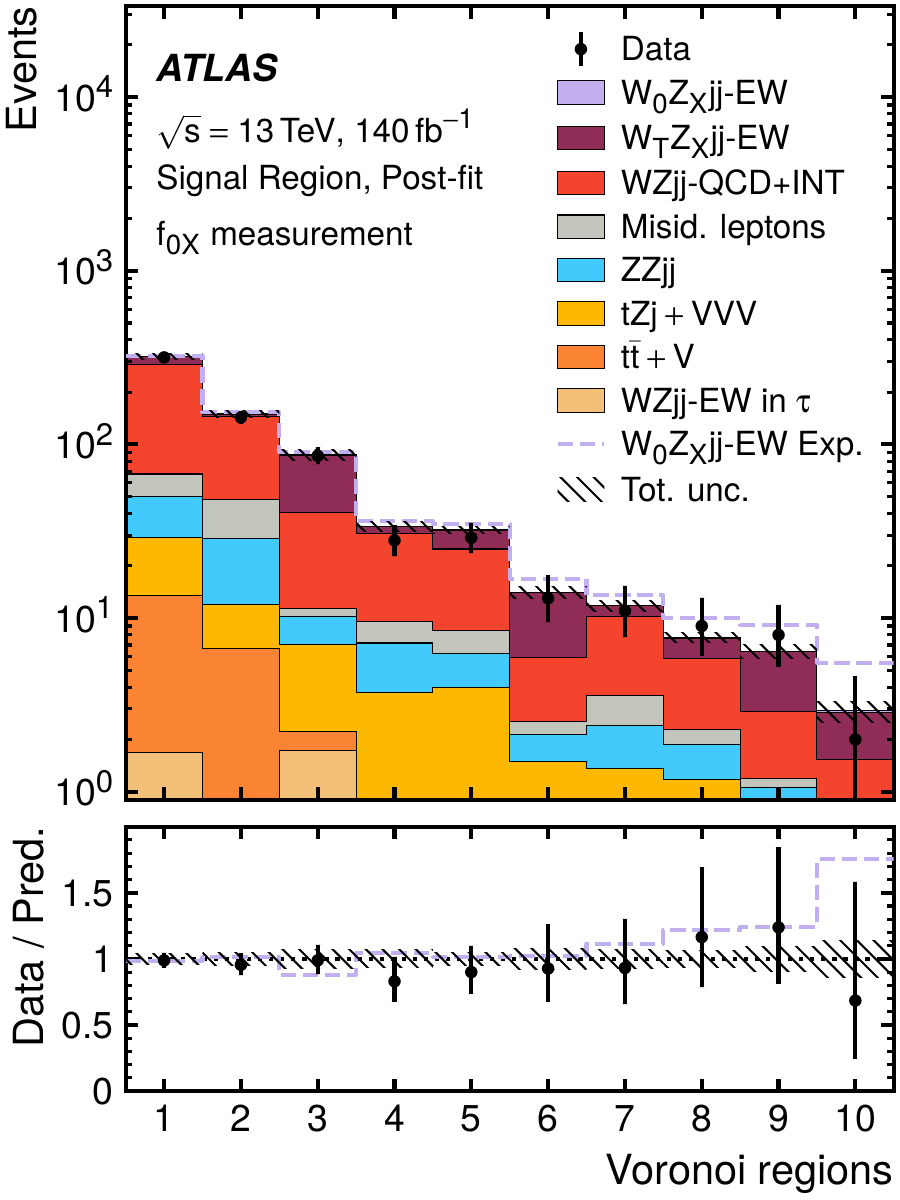}
}
\subfloat[]{
\includegraphics[width=0.33\textwidth]{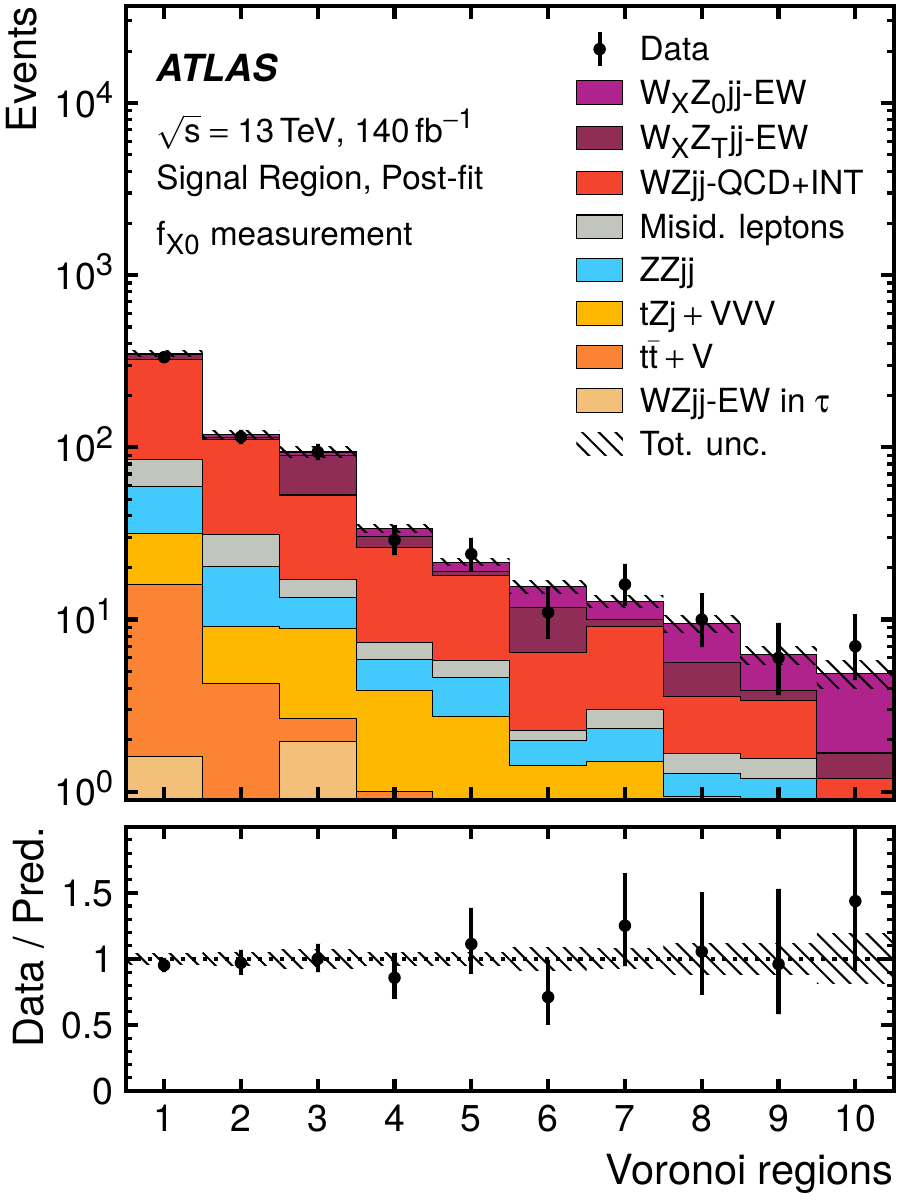}
}

\caption{Post-fit distributions for the Run 2 data sample of (a) \mjj in the \ZZCR control region, (b) the \CRbBDT score in the \bCR control region  and in the signal region the \fitDiscriName for (c) the \fzz measurement, (d) the \fzX measurement and (e) the \fXz measurement.
Signal and backgrounds are normalised to the expected number of events after the fit.
The sum of background events containing misidentified leptons is labelled ``Misid. leptons''.
The uncertainty band around the SM expectation includes all systematic uncertainties as obtained from the fit.
The fit for the \fXz measurement is used in (a), (b) and (e), while the fit for the \fzz and \fzX measurements are used in (c) and (d), respectively.
The dashed-line in (c) and (d) represent the sum of the post-fit SM predictions except for the polarised \wzjjew contributions that are scaled to their expected pre-fit values.
}
\label{fig:ScoreDistributions_Run2}
\end{figure}

\begin{figure}[t]
\centering

\subfloat[]{
\includegraphics[width=0.33\textwidth]{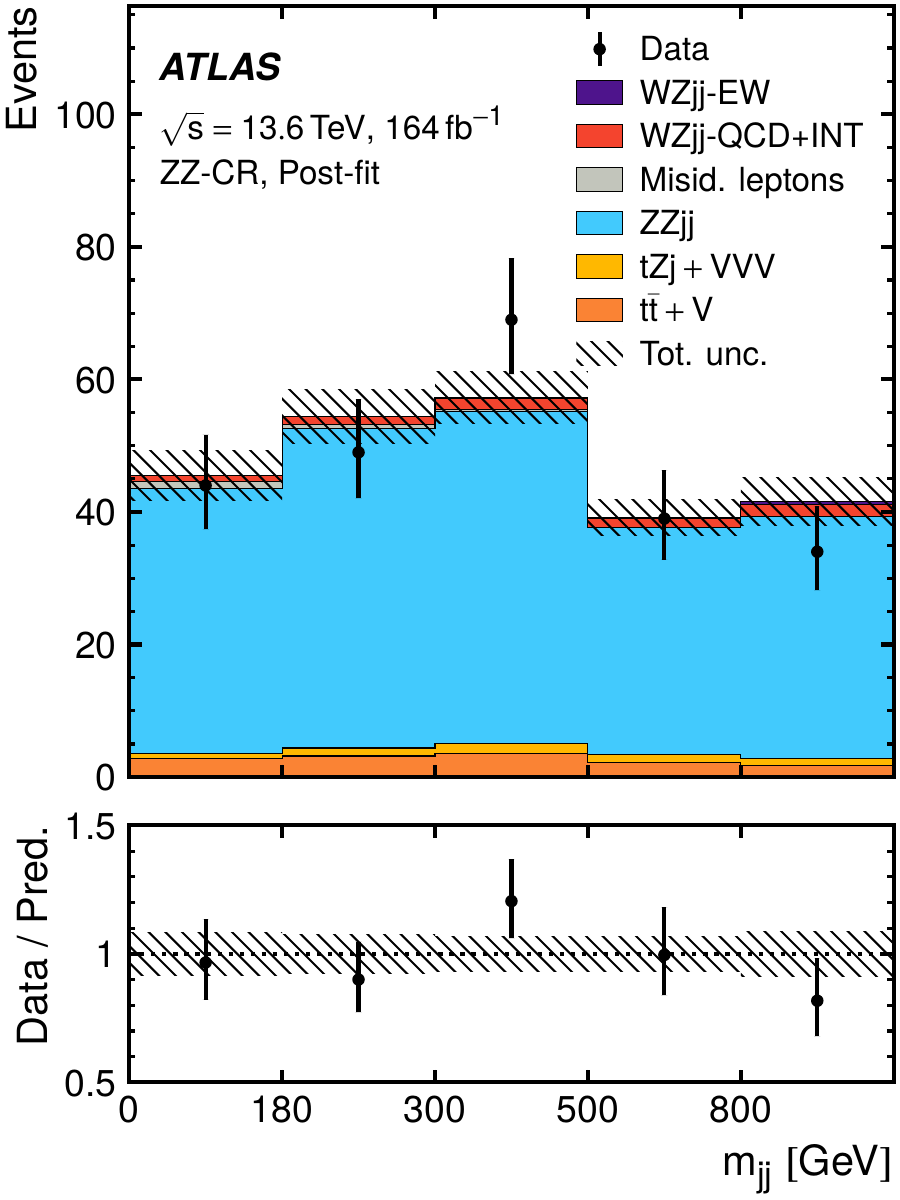}
}
\subfloat[]{
\includegraphics[width=0.33\textwidth]{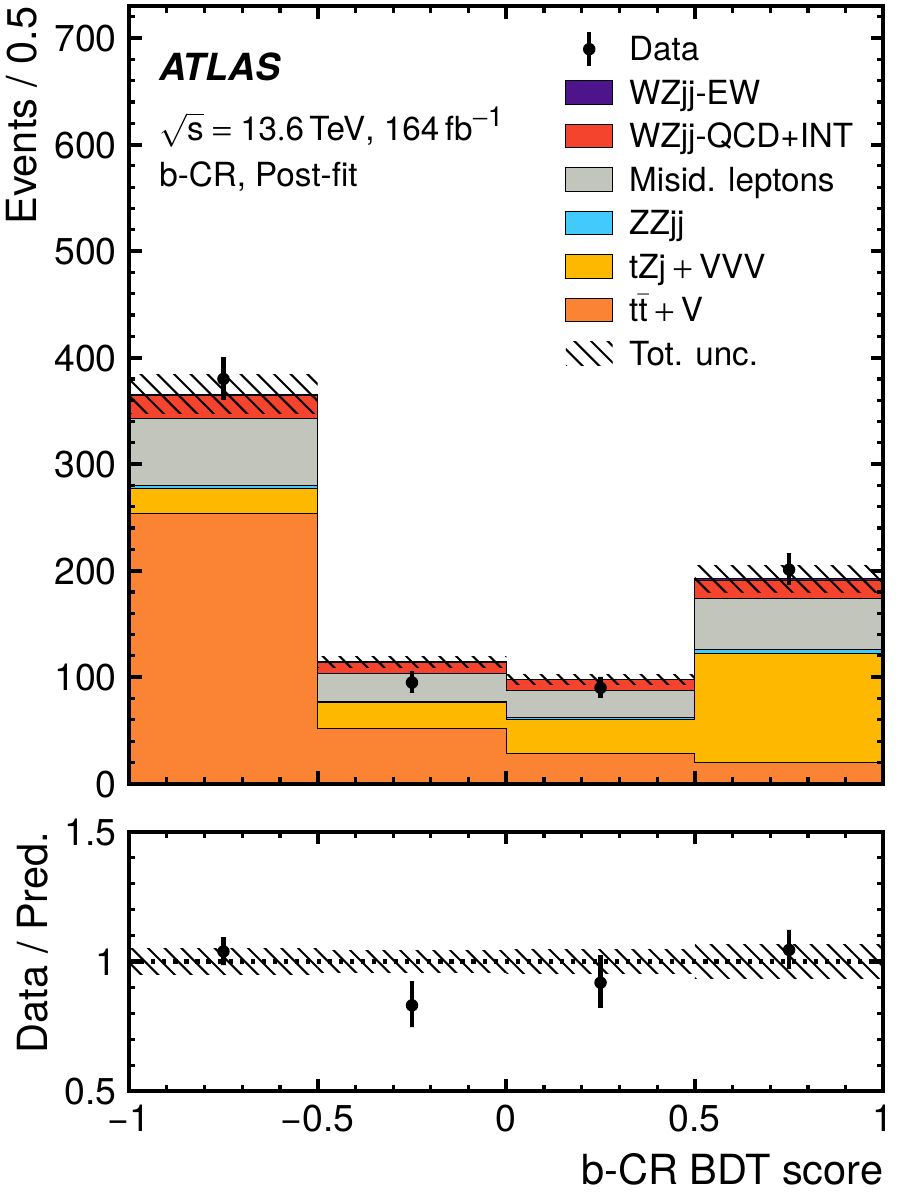}
}\\

\subfloat[]{
\includegraphics[width=0.33\textwidth]{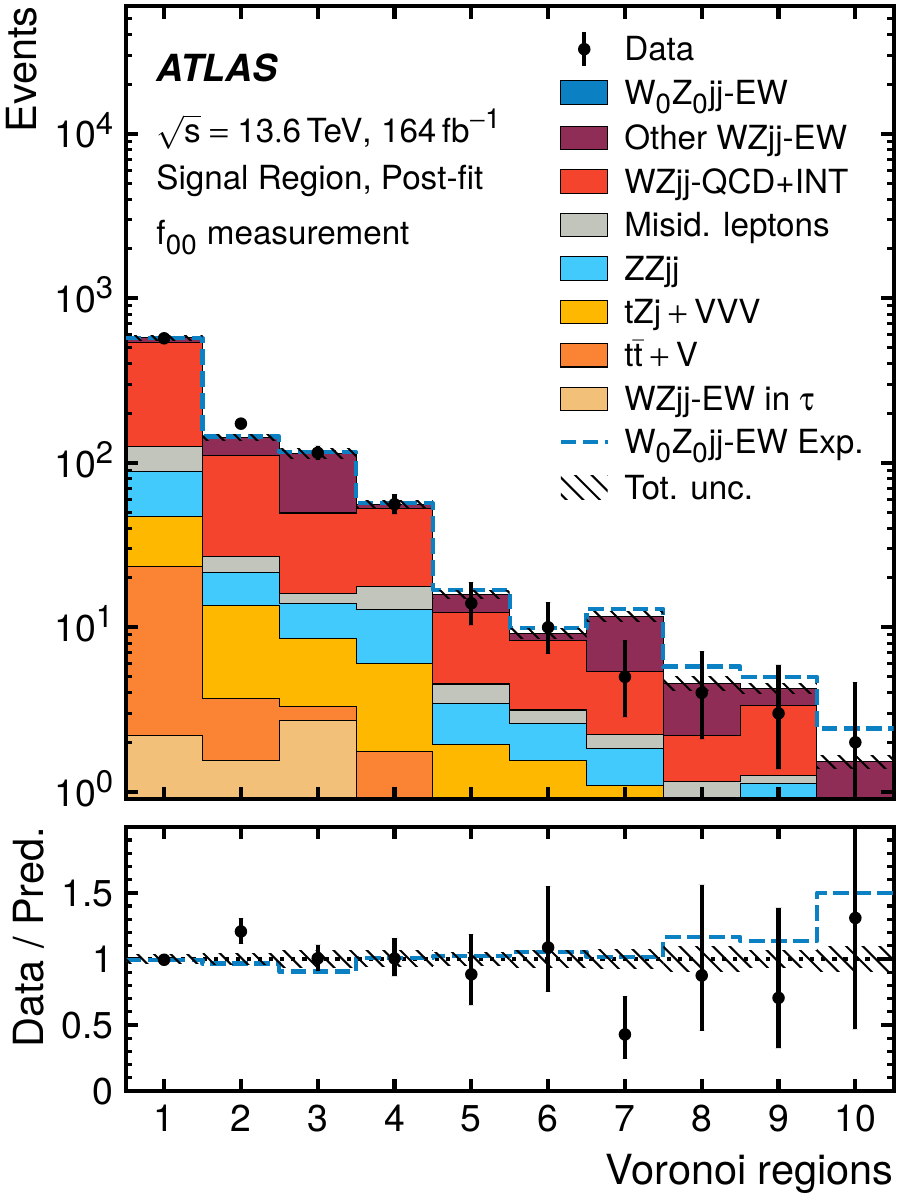}
}
\subfloat[]{
\includegraphics[width=0.33\textwidth]{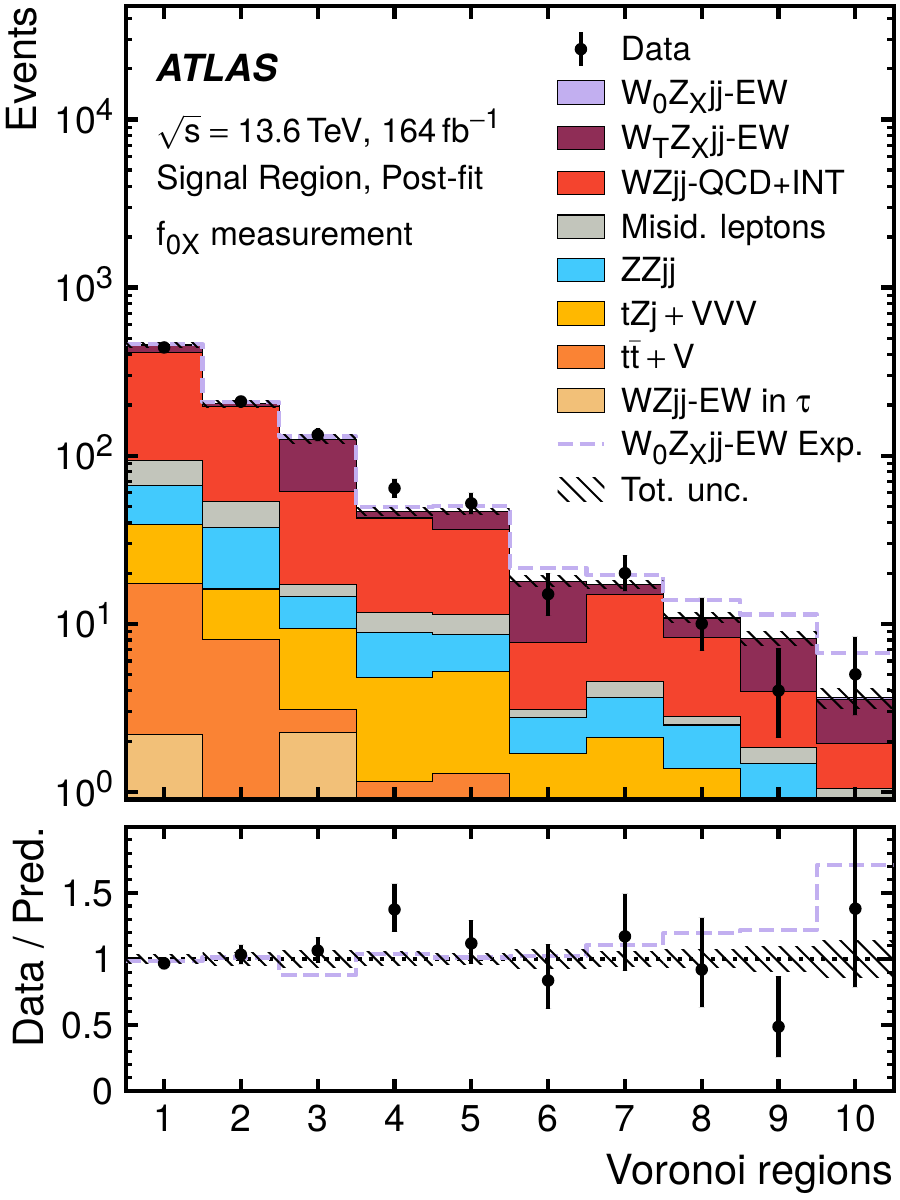}
}
\subfloat[]{
\includegraphics[width=0.33\textwidth]{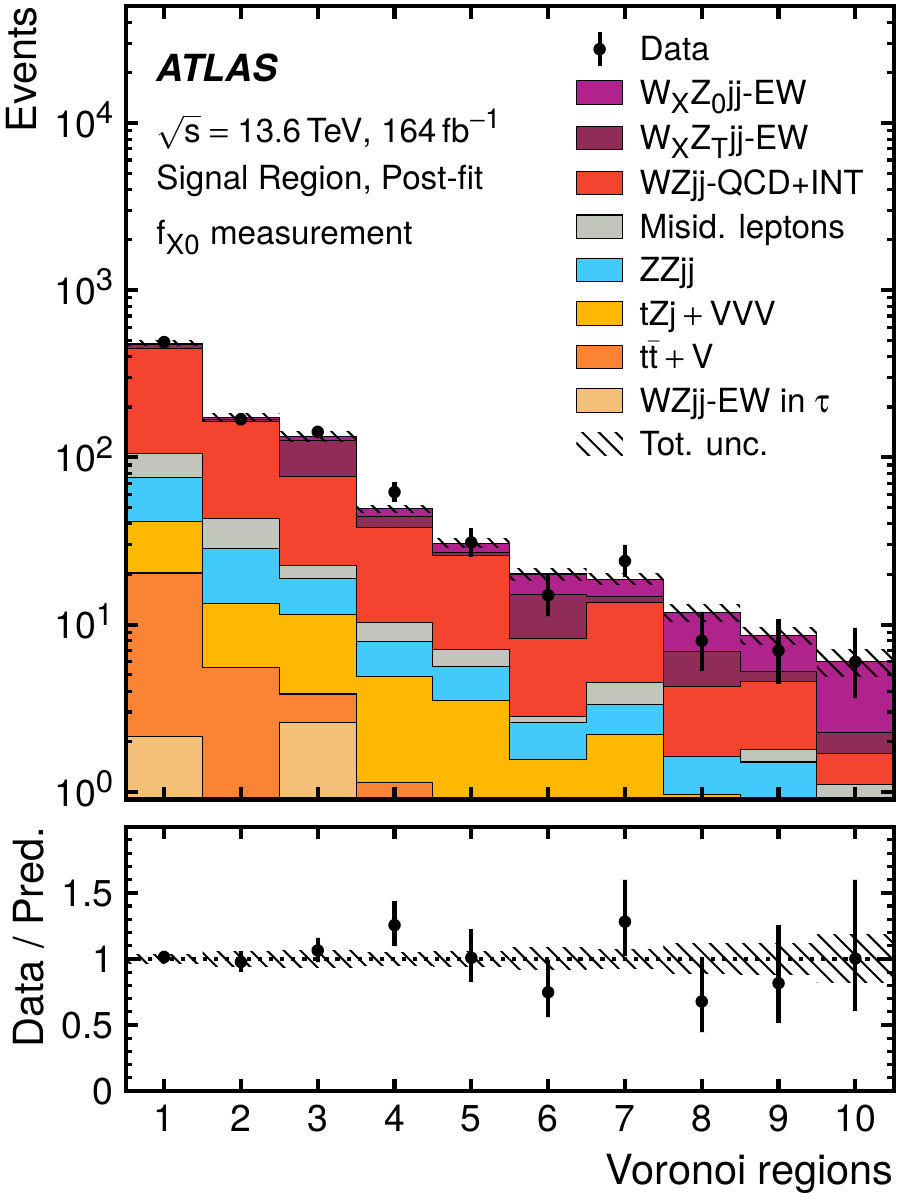}
}

\caption{Post-fit distributions for the Run 3 data sample of (a) \mjj in the \ZZCR control region, (b) the \CRbBDT score in the \bCR control region and in the signal region the \fitDiscriName for (c) the \fzz measurement, (d) the \fzX measurement and (e) the \fXz measurement.
Signal and backgrounds are normalised to the expected number of events after the fit.
The sum of background events containing misidentified leptons is labelled ``Misid. leptons''.
The uncertainty band around the SM expectation includes all systematic uncertainties as obtained from the fit.
The fit for the \fXz measurement is used in (a), (b) and (e), while the fit for the \fzz and \fzX measurements are used in (c) and (d), respectively.
The dashed-line in (c) and (d) represent the sum of the post-fit SM predictions except for the polarised \wzjjew contributions that are scaled to their expected pre-fit values.
}
\label{fig:ScoreDistributions_Run3}
\end{figure}

The hypotheses of having no events in each of the 00, 0X and X0 states are tested.
In \wzjjew events, the presence of a  $Z$ boson with a longitudinal polarisation independent of the polarisation state of the $W$ boson  (\fXz) is observed  with a significance of $4.0 \,\sigma$, compared with $2.9 \, \sigma$ expected.
No significant numbers of events from  \wzjjewzz or \wzjjewzx production are observed in data, while the expected significances for the presence of 00 and 0X events are of $1.3\sigma$ and $2.6\sigma$, respectively.
The observed $95\%$ confidence level (CL) upper limits on \fzz and \fzX are 0.12 and 0.19, respectively, while the expected upper limits were 0.21 and 0.46, respectively.
A value of $\muew = 1.06 \pm 0.18$ is obtained from the fit for \fXz.
Similar values of $\muew = 1.08 \pm 0.17$ and $\muew = 1.03 \pm 0.17$ are obtained in the two other fits for \fzz and \fzX, respectively.
The measurements are summarised in Table~\ref{tab:Results_PolRes} and compared with theoretical predictions from \Madgraph,  \sherpa 3.0 MC event generators and from \NLO EW fixed-order calculations using \MoCaNLO~\cite{Denner:2020wz}.
The fixed-order \MoCaNLO calculation is performed for \wpzjj production only and is  presented to assess a possible effect of NLO EW corrections on polarisation fractions.
Differences in polarisation fractions between \wpz and \wmz inclusive production were found to be at the percent level~\cite{Le:2022lrp,Le:2022ppa}, arising from quark flavours and PDF effects.
In \wzjjew production, owing to the fact that the \wz pair is mainly produced through vector boson scattering, these effects are expected to be strongly reduced.
Table~\ref{tab:Results:SystematicSummary} shows the main sources of uncertainty in the measurement of the \fXz polarisation fraction and of \muew.
The uncertainty related to data statistics is dominant in the measurement of  the \fXz polarisation fraction and determines its precision.
The values of \fzz, \fzX, \fXz, and \muew measured in \wzjjew events are represented in Figure~\ref{fig:Results:SummaryFigure}.
The probability to observe a value of $\fzX < 0.007$, as measured in data, under the SM hypothesis is $0.60\%$, corresponding to $2.5 \,\sigma$.
The correlations between the measured polarisation fractions and \muew values from each of the three fits are represented in Figure~\ref{fig:Results:Ellipses} and compared with MC predictions from \Madgraph, \Sherpa 3.0 and \NLO EW fixed-order calculations from \MoCaNLO.
Any effect of new physics enhancing or suppressing one polarisation state would manifest itself in these figures as a deviation in both the polarisation fraction and \muew with-respect to the SM prediction.
Fits are also performed individually on Run 2 and Run 3 data samples and yield consistent results for the measured \fzz, \fzX, \fXz, and \muew values, but with a lower sensitivity.

Performing the fit on the Run 3 data samples alone, the \wzjjew production cross-section at a centre-of-mass energy of $\sqrt{s} = 13.6$~\TeV, \sigewRthree, is measured\footnote{The \wzjjew production cross-section at $\sqrt{s} = 13$~\TeV was previously measured in Ref.~\cite{ATLAS:2024ini}.}.
A value of $\sigewRthree = 0.374 \; \pm 0.077 \,\mathrm{(stat.)} \; \pm 0.053 \,\mathrm{(syst.)} \; \pm 0.009 \,\mathrm{(lumi.)} \; \mathrm{fb}$ is obtained.
The precision on \sigewRthree does not depend on which polarisation fraction fit is performed, and the fit to \fXz is taken as the main cross-section result.
Values of \sigewRthree obtained from the fits for \fzX or \fzz agree within 10\%.
The corresponding prediction from \sherpa 3.0 is  $\sigewRthree = 0.373\; \pm 0.0003 \,\mathrm{(stat.)} \; \pm 0.008 \,\mathrm{(PDF)} \; ^{+0.065}_{-0.041} \,\mathrm{(scale)} \; \mathrm{fb}$, where the uncertainties correspond to statistical, PDF and QCD scale uncertainties, respectively, calculated as detailed in Section~\ref{sec:systematics}.

\begin{table}[!htbp]
\caption{
Measured polarisation fractions \fzz, \fzX, \fXz, and the \muew cross-section
in the fiducial phase space, for \wzjjew  events.
The \muew value obtained from the \fXz measurement is reported.
The total, statistical, systematic and luminosity uncertainties in the measurements are reported.
The measurements are compared with predictions from \Madgraph, \sherpa 3.0 and
from \NLO EW fixed-order calculations using \MoCaNLO~\cite{Denner:2020wz}.
The uncertainties in the \Madgraph and \sherpa predictions include statistical, PDF and QCD scale uncertainties; the uncertainties in the \NLO EW fixed-order prediction include QCD scale uncertainties.
As the \sherpa 3.0 prediction includes diagrams with up one additional parton at \LO in QCD, its uncertainty in \muew arising from higher order QCD corrections is larger than for the \madgraph prediction.
The calculation from \MoCaNLO is done for only $W^+Z$ events and therefore no inclusive \wzjjew cross-section is provided for it.
}
\label{tab:Results_PolRes}
\centering
\resizebox{0.98\textwidth}{!}{

%
%
%
%
%
%
%
%
%
%
%
%
%
%
%
%
%
%
%
\begin{tabular}{l c c c c}
\toprule
& Data & \multirow{2}{*}{\madgraph} & \multirow{2}{*}{\Sherpa~3.0} & \multirow{2}{*}{\MoCaNLO NLO EW} \\
& {\footnotesize $\pm$\,total ($\pm$\,stat., $\pm$\,syst., $\pm$\,lumi.)} & & \\
\midrule
$f_{\mathrm{00}}$ & \multicolumn{1}{l}{$0.000^{+0.053}_{-0.000}$~{\footnotesize $\left(^{+0.053}_{-0.000}\; ,\;<0.001,\;<0.0001 \right)$} } & $0.0902 \pm 0.0005$ & $0.087 \pm 0.001$ & $0.0902 \pm 0.0005$ \\[5pt]
$f_{\mathrm{0X}}$ & \multicolumn{1}{l}{ $0.007^{+0.094}_{-0.007}$~{\footnotesize $\left(^{+0.093}_{-0.000}\; ,\;^{+0.009}_{-0.007}\; ,\;^{+0.0001}_{-0.0005} \right)$} } & $0.2808 \pm 0.0002$ & $0.2793 \pm 0.0009$ & $0.2758 \pm 0.0008$ \\[5pt]
$f_{\mathrm{X0}}$ & \multicolumn{1}{l}{ $0.324^{+0.092}_{-0.085}$~{\footnotesize $\left(^{+0.091}_{-0.084}\; ,\;^{+0.017}_{-0.018}\; ,\;^{+0.0015}_{-0.0014}\right)$} }& $0.2758 \pm 0.0003$ & $0.274 \pm 0.001$ & $0.2745 \pm 0.0008$ \\
\midrule
\muew & $1.06 \pm 0.18$~{\footnotesize $(\pm 0.15,\;\pm 0.11,\;\pm 0.01)$} & 0.94 $^{+0.05}_{-0.04}$ & 1.00  $^{+0.2}_{-0.1}$ & {\text{---}} \\
\bottomrule
\end{tabular}

 %
}
\end{table}

\begin{figure}[!htbp]
\centering
\includegraphics[width=0.55\textwidth]{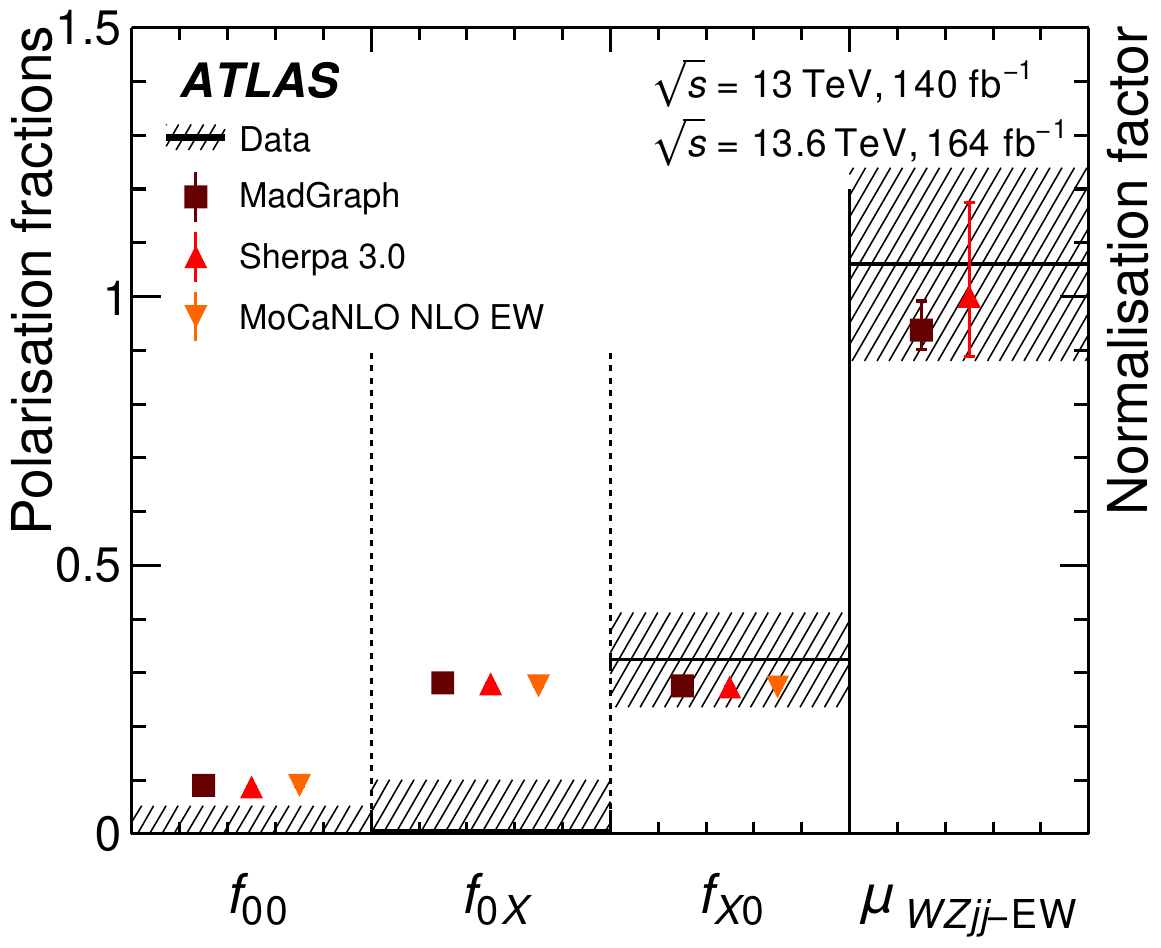}
\caption{Measured polarisation fractions \fzz, \fzX and \fXz of the $W$ and $Z$ bosons in \wzjjew events.
The normalisation factor \muew for the measured fiducial cross-section of \wzjjew production obtained from the \fXz measurement fit is also represented.
The measurements are compared with  predictions from \Madgraph (squares), \sherpa 3.0 (upward pointing triangles) and
from \NLO EW fixed-order calculations using \MoCaNLO~\cite{Denner:2020wz} (downward pointing triangles).
The normalisation factor \muew is relative to the \sherpa 3.0 prediction.
The calculations from \MoCaNLO are performed for only $W^+Z$ events and therefore no inclusive \wzjjew cross-section is shown.
}
\label{fig:Results:SummaryFigure}
\end{figure}

\begin{figure}[t]
\centering

\subfloat[]{
\includegraphics[width=0.45\textwidth]{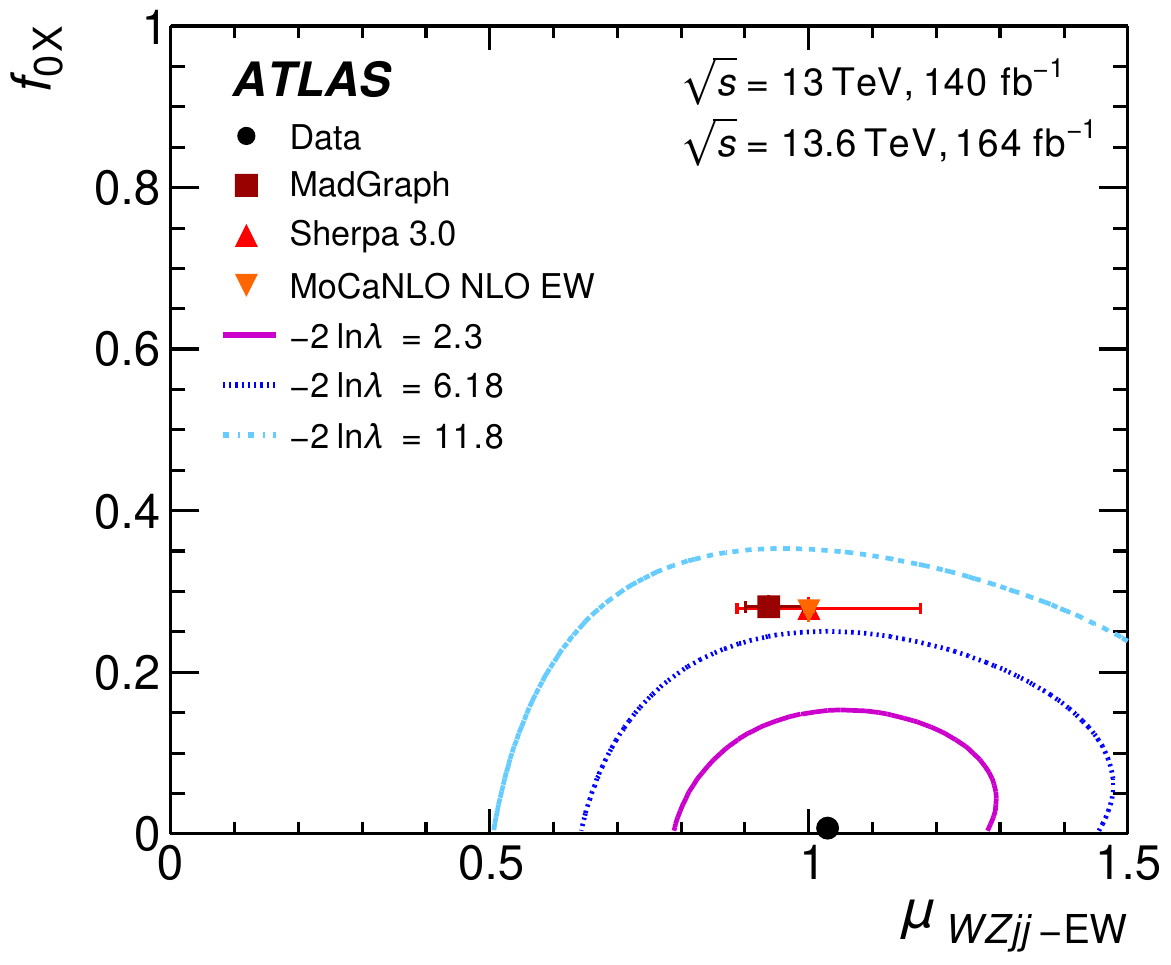}
}
\subfloat[]{
\includegraphics[width=0.45\textwidth]{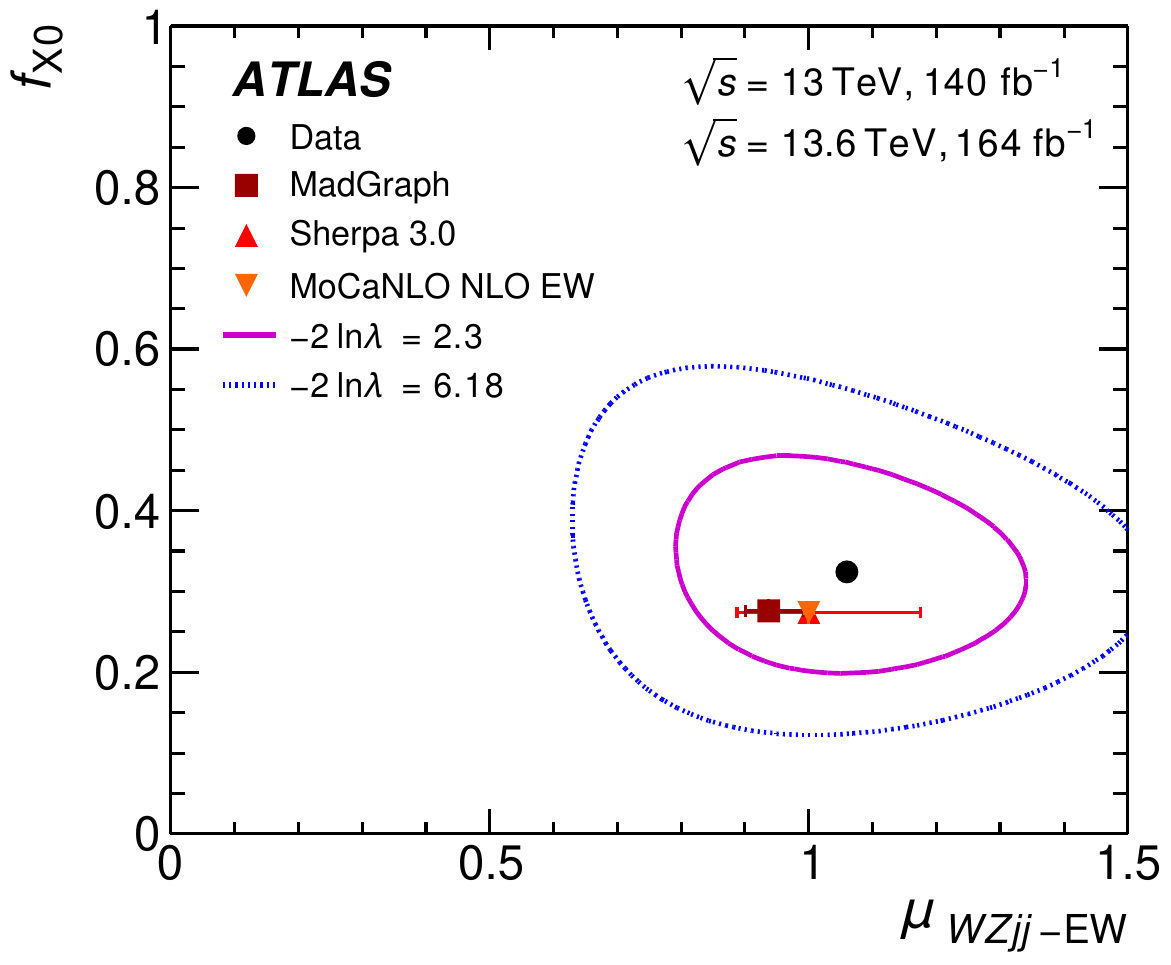}
}\\

\subfloat[]{
\includegraphics[width=0.45\textwidth]{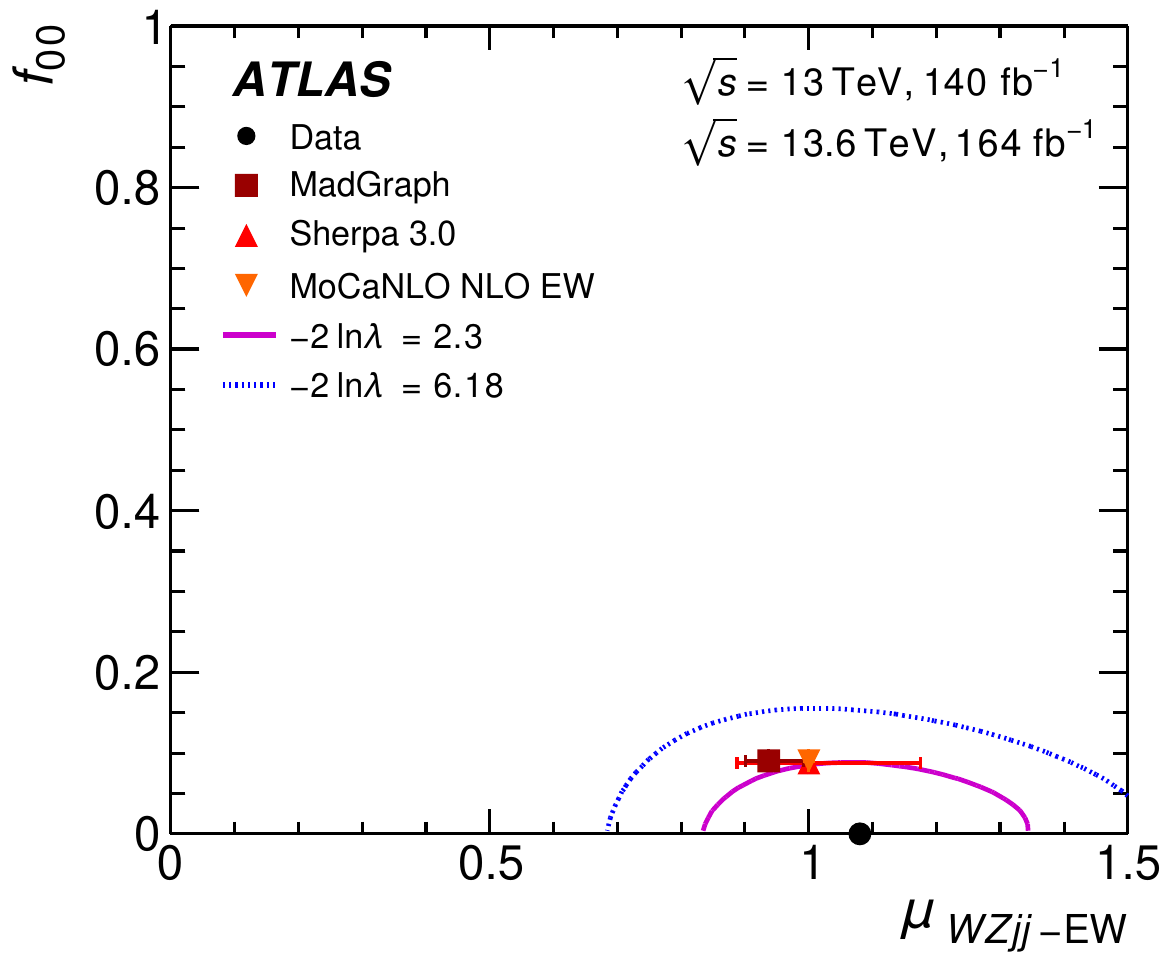}
}

\caption{The measured normalisation factor \muew and polarisation fractions (a) \fzX, (b) \fXz and (c) \fzz from the three independent fits. The measurements are compared with MC predictions from \Madgraph (squares), \Sherpa 3.0 (upward pointing triangles) and \NLO EW fixed-order calculations from \MoCaNLO (downward pointing triangles).
The full and dashed ellipses around the data points correspond to contours were the likelihood ratio $-2 \ln \lambda$ is equal to 2.3 and 6.18, respectively.
In (a) the contour corresponding to $-2 \ln \lambda = 11.8$ is also represented by a dotted-dashed ellipse.
The calculations from \MoCaNLO are performed for only $W^+Z$ events so its cross-section prediction is set to that of \sherpa ($\muew = 1$).
}
\label{fig:Results:Ellipses}
\end{figure}


%
%
\FloatBarrier

\section{Conclusion}
\label{sec:conclusion}
%

Measurements of gauge boson polarisation states from \wz electroweak production  in $\sqrt{s} = 13$~\TeV\ and  $\sqrt{s} = 13.6$~\TeV\ $pp$ collisions at the LHC are presented.
The data analysed were collected with the ATLAS detector during the years $2015$ to $2024$ and correspond to an integrated luminosity of $304~\ifb$.
The measurements use leptonic decay modes of the gauge bosons to electrons or muons and are performed in a fiducial phase space closely matching the detector acceptance.

Using neural-network-based analysis techniques,  the fraction of \wzjjew events with a $Z$ boson longitudinally polarised is measured for the first time with a significance of 4.0 standard deviations, for an expected significance of 2.9 standard deviations.
Integrated over the fiducial region, the fraction is measured to be $f_{\mathrm{X0}} = 0.32 \pm 0.09$, in agreement with the Standard Model prediction of $f_{\mathrm{X0}} = 0.2745 \pm 0.0008$, at a precision of next-to-leading-order in electroweak corrections.
No evidence for the production of longitudinally polarised $W$ bosons is observed in the data.
Observed upper limits at 95\% confidence level of 0.19 and 0.12 are set on the fraction of events with a $W$ boson longitudinally polarised and on the fraction of events with the two bosons longitudinally polarised, respectively.
Corresponding expected upper limits are 0.46 and 0.21, respectively.
The non-observation of longitudinally polarised $W$ bosons independently of the polarisation of the $Z$ boson is within 2.5 standard deviations from the SM expectation.

Furthermore, the inclusive cross-section of \wzjjew production at $\sqrt{s} = 13.6$~\TeV in the fiducial region for leptonic decay modes is measured to be $\sigma_{WZjj\mathrm{-EW} \rightarrow \ell^{'} \nu \ell \ell jj} = 0.374 \; \pm  0.077 \,(\mathrm{stat.}) \; \pm 0.053  \,(\mathrm{syst.}) \; \pm 0.009 \,(\mathrm{lumi.})  \; \mathrm{fb}$, where $\ell$ and $\ell^{'}$ are either an electron or a muon.
The measurement is in agreement with the SM prediction from the \sherpa event generator of  $\sigewRthree = 0.373\;  ^{+0.065}_{-0.042} \; \mathrm{fb}$.


%
%
%
%
%
\section*{Acknowledgements}
%

%

%
%

%
%

We thank CERN for the very successful operation of the LHC and its injectors, as well as the support staff at
CERN and at our institutions worldwide without whom ATLAS could not be operated efficiently.

The crucial computing support from all WLCG partners is acknowledged gratefully, in particular from CERN, the ATLAS Tier-1 facilities at TRIUMF/SFU (Canada), NDGF (Denmark, Norway, Sweden), CC-IN2P3 (France), KIT/GridKA (Germany), INFN-CNAF (Italy), NL-T1 (Netherlands), PIC (Spain), RAL (UK) and BNL (USA), the Tier-2 facilities worldwide and large non-WLCG resource providers. Major contributors of computing resources are listed in Ref.~\cite{ATL-SOFT-PUB-2026-001}.

We gratefully acknowledge the support of ANPCyT, Argentina; YerPhI, Armenia; ARC, Australia; BMWFW and FWF, Austria; ANAS, Azerbaijan; CNPq and FAPESP, Brazil; NSERC, NRC and CFI, Canada; CERN; ANID, Chile; CAS, MOST, NSFC and FRFCU, China; Minciencias, Colombia; MEYS CR, Czech Republic; DNRF and DNSRC, Denmark; IN2P3-CNRS and CEA-DRF/IRFU, France; SRNSFG, Georgia; BMFTR, HGF and MPG, Germany; GSRI, Greece; RGC and Hong Kong SAR, China; ICHEP and Academy of Sciences and Humanities, Israel; INFN, Italy; MEXT and JSPS, Japan; CNRST, Morocco; NWO, Netherlands; RCN, Norway; MNiSW, Poland; FCT, Portugal; MNE/IFA, Romania; MSTDI, Serbia; MSSR, Slovakia; ARIS and MVZI, Slovenia; DSI/NRF, South Africa; MICIU/AEI, Spain; SRC and Wallenberg Foundation, Sweden; SERI, SNSF and Cantons of Bern and Geneva, Switzerland; NSTC, Taipei; TENMAK, T\"urkiye; STFC/UKRI, United Kingdom; DOE and NSF, United States of America.

Individual groups and members have received support from BCKDF, CANARIE, CRC and DRAC, Canada; CERN-CZ, FORTE and PRIMUS, Czech Republic; COST, ERC, ERDF, Horizon 2020 and Marie Sk{\l}odowska-Curie Actions, European Union; Investissements d'Avenir Labex, Investissements d'Avenir Idex and ANR, France; DFG and AvH Foundation, Germany; Herakleitos, Thales and Aristeia programmes co-financed by EU-ESF and the Greek NSRF, Greece; BSF-NSF and MINERVA, Israel; NCN and NAWA, Poland; La Caixa Banking Foundation, CERCA and AGAUR programs from Generalitat de Catalunya and PROMETEO and GenT Programmes Generalitat Valenciana, Spain; G\"{o}ran Gustafssons Stiftelse, Sweden; The Royal Society and Leverhulme Trust, United Kingdom; Eric and Wendy Schmidt Fund for Strategic Innovation, United States of America.

In addition, individual members wish to acknowledge support from Chile: Agencia Nacional de Investigaci\'on y Desarrollo (ANID FONDECYT reg. 1230987, FONDECYT 1230812, FONDECYT 1240864, Fondecyt 3240661, Fondecyt Regular 1240721); China: Fundamental Research Funds for the Central Universities (010-63263105), Chinese Ministry of Science and Technology (MOST-2023YFA1605700, MOST-2023YFA1609300), National Natural Science Foundation of China (NSFC 12275265, NSFC-W2543005); Czech Republic: Czech Science Foundation (GACR - 24-11373S), Ministry of Education Youth and Sports (ERC-CZ-LL2327, FORTE CZ.02.01.01/00/22\_008/0004632); EU: H2020 European Research Council (ERC - 101002463); European Union: European Research Council (BARD No. 101116429, ERC - 101219398, ERC 101089007), European Regional Development Fund (HE COFUND GA No.101081355, ERDF), Marie Sklodowska-Curie Actions (GAP-101168829, GAP-101149251); France: Agence Nationale de la Recherche (ANR-22-EDIR-0002, ANR-24-CE31-0504-01); Germany: Deutsche Forschungsgemeinschaft (DFG - 469666862); China: Research Grants Council (GRF); Italy: Istituto Nazionale di Fisica Nucleare (LHC-MIUR - 28003/2025), Ministero dell'Università e della Ricerca (NextGenEU I53D23000820006 M4C2.1.1, SOE2024\_0000023); Japan: Japan Society for the Promotion of Science (JSPS KAKENHI  JP25H0063, JSPS KAKENHI JP22H04944, JSPS KAKENHI JP24K23939, JSPS KAKENHI JP24KK0251, JSPS KAKENHI JP25H00650, JSPS KAKENHI JP25H01291, JSPS KAKENHI JP25K01011, JSPS KAKENHI JP25K01023, JSPS KAKENHI JP25KK0047); Poland: Polish National Science Centre (NCN 2021/42/E/ST2/00350, NCN OPUS 2023/51/B/ST2/02507, NCN OPUS nr 2022/47/B/ST2/03059, NCN UMO-2019/34/E/ST2/00393, UMO-2022/47/O/ST2/00148, UMO-2023/49/B/ST2/04085, UMO-2023/51/B/ST2/00920, UMO-2024/53/N/ST2/00869); Spain: Agència de Gestió d'Ajuts Universitaris i de Recerca. (AGAUR - 2023 BP 00141), Ministry of Science and Innovation (RYC2019-028510-I, RYC2020-030254-I, RYC2021-031273-I, RYC2022-038164-I), Ministerio de Ciencia, Innovación y Universidades/Agencia Estatal de Investigaci\'on (EU NextGenerationEU (PRTR-C17.I1), PID2022-142604OB-C22); Sweden: Carl Trygger Foundation (Carl Trygger Foundation CTS 22:2312), Swedish Research Council (Swedish Research Council 2023-04654, VR 2021-03651, VR 2022-03845, VR 2022-04683, VR 2023-03403, VR 2024-05451, VR 2025-05940), Knut and Alice Wallenberg Foundation (KAW 2023.0366); United Kingdom: The Binks Trust, Royal Society (NIF-R1-231091); United States of America: U.S. Department of Energy (ECA DE-AC02-76SF00515), John Templeton Foundation (John Templeton Foundation 63206), Neubauer Family Foundation.

%
%


%
%
%
%
%
%

%
%

%
%
\printbibliography
\clearpage
 
\begin{flushleft}
\hypersetup{urlcolor=black}
{\Large The ATLAS Collaboration}

\bigskip

\AtlasOrcid[0000-0002-6665-4934]{G.~Aad}$^\textrm{\scriptsize 102}$,
\AtlasOrcid[0000-0001-7616-1554]{E.~Aakvaag}$^\textrm{\scriptsize 17}$,
\AtlasOrcid[0000-0002-5888-2734]{B.~Abbott}$^\textrm{\scriptsize 121}$,
\AtlasOrcid[0000-0002-0287-5869]{S.~Abdelhameed}$^\textrm{\scriptsize 83b}$,
\AtlasOrcid[0000-0002-1002-1652]{K.~Abeling}$^\textrm{\scriptsize 54}$,
\AtlasOrcid[0000-0001-5763-2760]{N.J.~Abicht}$^\textrm{\scriptsize 48}$,
\AtlasOrcid[0000-0002-8496-9294]{S.H.~Abidi}$^\textrm{\scriptsize 30}$,
\AtlasOrcid[0009-0003-6578-220X]{M.~Aboelela}$^\textrm{\scriptsize 44}$,
\AtlasOrcid[0000-0002-9987-2292]{A.~Aboulhorma}$^\textrm{\scriptsize 36e}$,
\AtlasOrcid[0000-0001-5329-6640]{H.~Abramowicz}$^\textrm{\scriptsize 154}$
% ...
\bigskip
\\

% ...
$^{17}$Department for Physics and Technology, University of Bergen, Bergen; Norway.\\
% ...
$^{30}$Physics Department, Brookhaven National Laboratory, Upton NY; United States of America.\\
% ...
$^{36}$$^{(a)}$Facult\'e des Sciences Ain Chock, Universit\'e Hassan II de Casablanca;$^{(b)}$Facult\'{e} des Sciences, Universit\'{e} Ibn-Tofail, K\'{e}nitra;$^{(c)}$Facult\'e des Sciences Semlalia, Universit\'e Cadi Ayyad, LPHEA-Marrakech;$^{(d)}$LPMR, Facult\'e des Sciences, Universit\'e Mohamed Premier, Oujda;$^{(e)}$Facult\'e des sciences, Universit\'e Mohammed V, Rabat;$^{(f)}$Institute of Applied Physics, Mohammed VI Polytechnic University, Ben Guerir; Morocco.\\
% ...
$^{44}$Physics Department, Southern Methodist University, Dallas TX; United States of America.\\
% ...
$^{48}$Fakult\"{a}t Physik, Technische Universit{\"a}t Dortmund, Dortmund; Germany.\\
% ...
$^{54}$II. Physikalisches Institut, Georg-August-Universit\"{a}t G\"ottingen, G\"ottingen; Germany.\\
% ...
$^{83}$$^{(a)}$Khalifa University of Science and Technology, Abu Dhabi;$^{(b)}$New York University Abu Dhabi, Abu Dhabi;$^{(c)}$United Arab Emirates University, Al Ain;$^{(d)}$University of Sharjah, Sharjah; United Arab Emirates.\\
% ...
$^{102}$CPPM, Aix-Marseille Universit\'e, CNRS/IN2P3, Marseille; France.\\
% ...
$^{121}$Homer L. Dodge Department of Physics and Astronomy, University of Oklahoma, Norman OK; United States of America.\\
% ...
$^{154}$Raymond and Beverly Sackler School of Physics and Astronomy, Tel Aviv University, Tel Aviv; Israel.\\
% ...
$^{a}$ Also at Affiliated with an institute formerly covered by a cooperation agreement with CERN.\\
$^{b}$ Also at An-Najah National University, Nablus; Palestine.\\
$^{c}$ Also at Borough of Manhattan Community College, City University of New York, New York NY; United States of America.\\
$^{d}$ Also at Center for Interdisciplinary Research and Innovation (CIRI-AUTH), Thessaloniki; Greece.\\
$^{e}$ Also at Centre of Physics of the Universities of Minho and Porto (CF-UM-UP); Portugal.\\
$^{f}$ Also at CERN, Geneva; Switzerland.\\
% ...

\end{flushleft}


%

%
%
\clearpage
\newpage

\end{document}